\documentclass[aps,prl,superscriptaddress,twocolumn,10pt,longbibliography]{revtex4-1}

\usepackage{graphicx}
\usepackage{amsmath}
\usepackage{amssymb}
\usepackage{dsfont}
\usepackage{xspace}
\usepackage{color}
\usepackage[usenames,dvipsnames]{xcolor}
\usepackage[colorlinks=true,linkcolor=blue,urlcolor=blue,citecolor=blue]{hyperref}
\usepackage[normalem]{ulem}
\usepackage{times}

\DeclareMathOperator{\re}{\mbox{Re}}
\DeclareMathOperator{\im}{\mbox{Im}}
\DeclareMathOperator{\erfc}{{erfc}}

\newcommand{\nag}{{\phantom{\dagger}}}

\newcommand{\eqw}[1]{(\ref{#1})}
\newcommand{\eq}[1]{Eq.~(\ref{#1})}

\newcommand{\eqs}[1]{Eq.~({#1})}

\newcommand{\fig}[1]{Fig.\thinspace{}\ref{#1}}

\newcommand{\fc}[1]{({#1})}
\newcommand{\figc}[2]{Fig.\thinspace{}\ref{#1}\thinspace{}\fc{#2}}

\def\bra#1{\mathinner{\langle{#1}|}}
\def\ket#1{\mathinner{|{#1}\rangle}}

\def\beq{\begin{equation}}
\def\eeq{\end{equation}}
\def\bea{\begin{eqnarray}}
\def\eea{\end{eqnarray}}

\begin{document}

\title{Noise-induced subdiffusion in strongly localized quantum systems}

\author{Sarang Gopalakrishnan}
\affiliation{Department of Engineering Science and Physics, CUNY College of Staten Island, Staten Island, NY 10314, USA}
\affiliation{Department of Physics and Walter Burke Institute, California Institute of Technology, Pasadena, CA 91125, USA}%

\author{K. Ranjibul Islam}
\affiliation{Department of Physics and Astronomy, Texas A\&M University, College Station, TX 77843, USA}
\affiliation{Indian Institute of Science Education and Research-Kolkata, Mohanpur, Nadia-741246, India}

\author{Michael Knap}%
\affiliation{Department of Physics, Walter Schottky Institute, and Institute for Advanced Study, Technical University of Munich, 85748 Garching, Germany}%

\date{\today}

\begin{abstract}

We consider the dynamics of strongly localized systems subject to dephasing noise with arbitrary correlation time. Although noise inevitably induces delocalization, transport in the noise-induced delocalized phase is subdiffusive in a parametrically large intermediate-time window. We argue for this intermediate-time subdiffusive regime both analytically and using numerical simulations on single-particle localized systems. Furthermore, we show that normal diffusion is restored in the long-time limit, through processes analogous to variable-range hopping. With numerical simulations based on Lanczos exact diagonalization, we demonstrate that our qualitative conclusions are also valid for interacting systems in the many-body localized phase.

\end{abstract}

\maketitle

The effects of disorder on quantum transport and dynamics have been a topic of longstanding interest~\cite{lee_RMP, kramer_RMP}. Both noninteracting~\cite{anderson_absence_1958} and interacting~\cite{fleishman, basko_metalinsulator_2006, gornyi_interacting_2005, nandkishore_mbl_2015, altman_universal_2015, Schreiber15, kondov_Disorder_2015, smith2015, Bordia16, choi_exploring_2016, bordia_periodically_2016} systems of electrons in a random potential can get ``localized'' by disorder, causing their d.c.~conductivity to vanish in the limit of a fully isolated system. Isolated localized systems not only have vanishing transport coefficients, but also fail to reach thermal equilibrium starting from generic initial conditions~\cite{nandkishore_mbl_2015}.
Yet, in any practical situation the system of interest is coupled to a thermalizing environment, which restores equilibrium and transport. The nature of equilibration in the presence of a bath has been a topic of recent interest~\cite{basko_expt, Nandkishore14, gopalakrishnan2014mean, johri2015many, huse2015localized, hyatt_many-body_2017, parameswaran_spin-catalyzed_2017, fischer2016dynamics, levi2016robustness, medvedyeva2016influence, everest_role_2017, nandkishore_general}; however the implications for \emph{transport} have not yet been investigated in general (but see Refs.~\cite{basko_expt, parameswaran_spin-catalyzed_2017}). 
\begin{figure}
  \centering 
  \includegraphics[width=.48\textwidth]{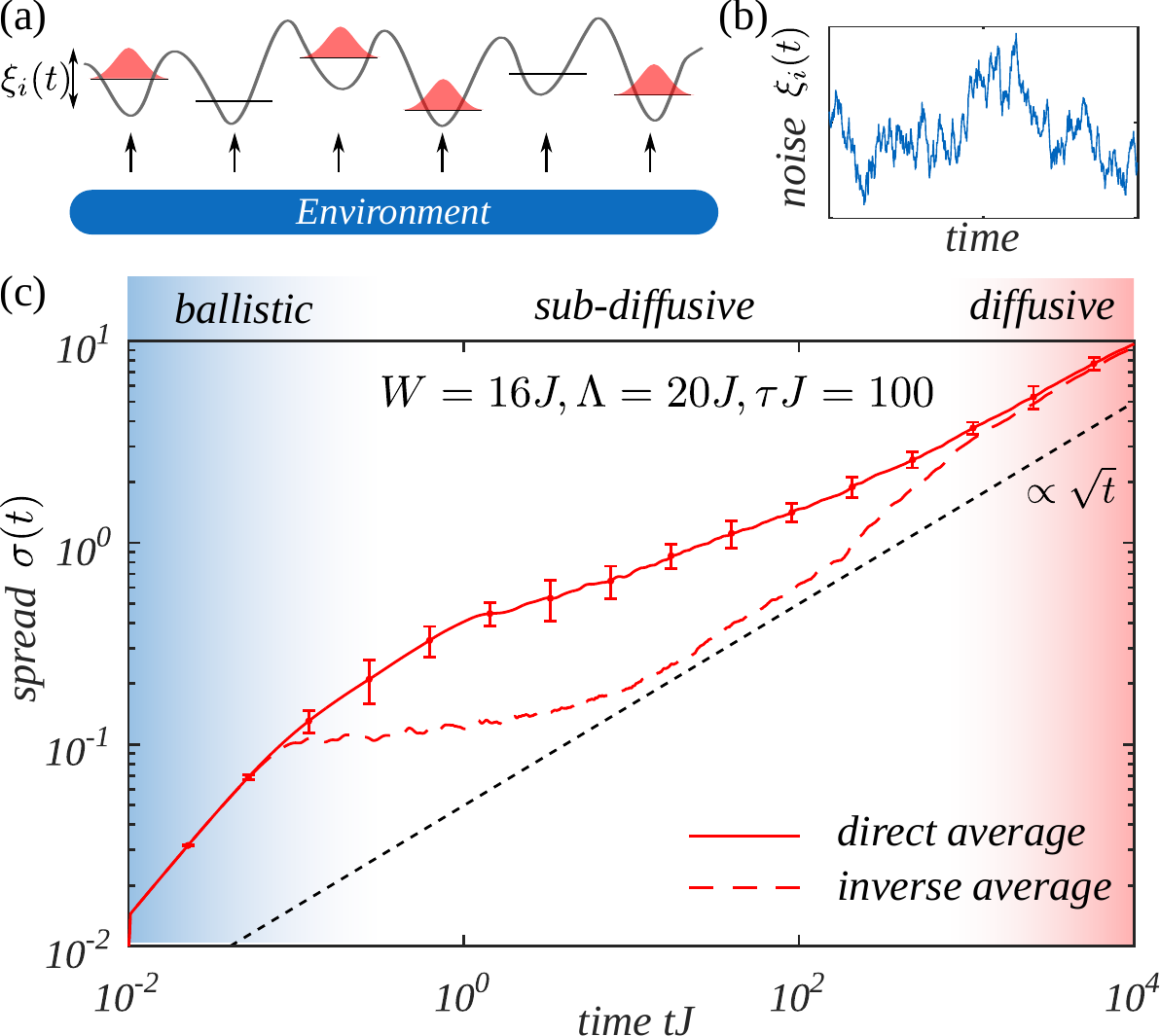}
  \caption{ \textbf{Noise-induced delocalization.} \fc{a} We consider strongly-localized fermions in a random potential that are weakly coupled to an environment. \fc{b} In that limit, the environment can be modeled by classical noise $\xi_i(t)$ that couples locally to the density. \fc{c} We study the noise-induced transport, by preparing the system in a wavepacket and computing its spread $\sigma(t)$ in time. For finite coupling to the environment three regimes can be distinguished: (i) a short-time ballistic expansion $\sigma(t) \sim t$, (ii) a parametrically large regime of subdiffusive transport $\sigma(t) \sim t^\beta$ with a continuously increasing power $\beta$ that approaches (iii) a diffusive regime $\sigma(t) \sim \sqrt{t}$ at late times. Numerical data taken as direct (solid) and inverse (dashed) average of the spread over individual realizations are shown for disorder  $W=16J$, noise strength $\Lambda=20J$, and noise correlation time $\tau J=100$. 
  } 
  \label{fig:schematic}
\end{figure}

One reason to expect unusual transport properties in \emph{imperfectly} isolated localized states is that even \emph{perfectly} isolated localized states exhibit a broad distribution of timescales. This feature was recently noticed as a property of dynamics near the many-body localization (MBL) transition, where transport has been found to be anomalous~\cite{BarLev_Absence_2015, agarwal_anomalous_2015, Vosk_Theory_2015, Potter_Universal_2015, luitz2016, gopalakrishnan_griffiths_2016, agarwal_rr_16, lueschencrit16, bordia2d17}. However, properties such as overlap integrals between localized orbitals also exhibit broad distributions deep in the localized phase~\cite{Gopalakrishnan15,pekker_2016,serbyn_interferometric_2014}. Consequently, the inter-orbital hopping rates induced by the bath are broadly distributed~\cite{fischer2016dynamics, levi2016robustness, medvedyeva2016influence}. One might expect such broad distributions to have anomalous transport signatures, particularly in one-dimensional systems, where single weak links can blockade transport.

In the present work, we explore this question, for localized systems coupled to generic non-Markovian dephasing noise. The Markovian limit was previously considered in Refs.~\cite{fischer2016dynamics, levi2016robustness, medvedyeva2016influence}; these works noted a broad distribution of relaxation times, leading to stretched-exponential decay of the ``contrast'' (as measured in Ref.~\cite{Schreiber15}). We find that slowly fluctuating noise can have even more dramatic effects: for strong disorder and slowly fluctuating noise, we find a large intermediate time window in which the system exhibits anomalous diffusion (\fig{fig:schematic}). This anomalous regime vanishes in the limit of Markovian noise, and also crosses over to diffusion in the long-time limit. The existence of a subdiffusive regime is notable because we do not explicitly \emph{introduce} any broad distributions, as opposed to the cases in Ref.~\cite{bouchaud_anomalous_1990}. Rather, a broad distribution of hopping rates \emph{emerges} from the interplay between disorder, quantum localization effects, and noise, as we discuss below. Furthermore, unlike the subdiffusive regime prefiguring the MBL transition~\cite{agarwal_anomalous_2015}, the phenomenon we discuss here is present in the noninteracting limit. 

Our focus is on free-fermion systems coupled to classical colored noise, as relatively large systems are accessible in numerical simulations for this case. As we discuss and substantiate with numerical simulations, however, our qualitative conclusions can also be adapted to \emph{interacting} systems in the MBL phase. Moreover, our model can be extended from classical noise to quantum dephasing, using a mapping between these two processes~\cite{crow_2014}.

\begin{figure*}
  \centering 
  \includegraphics[width=.98\textwidth]{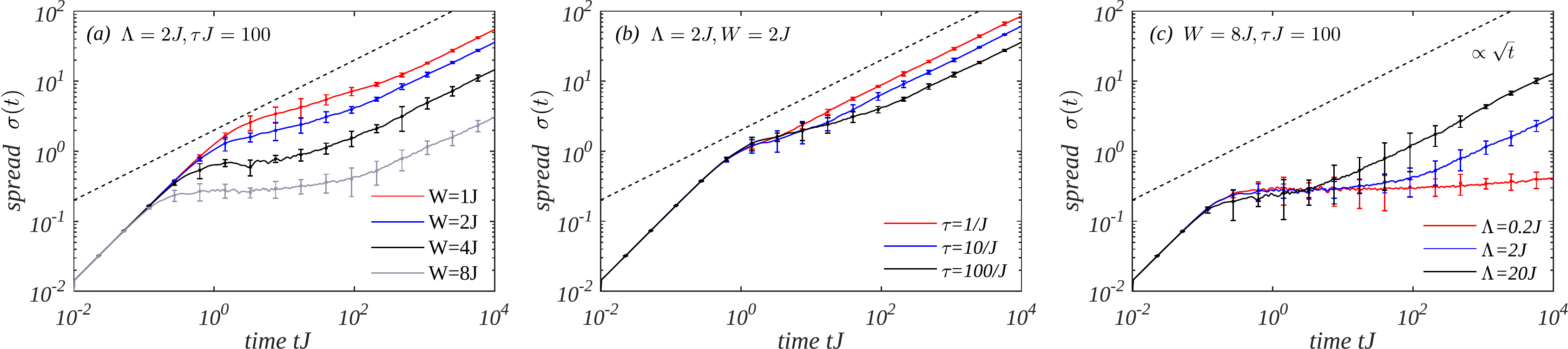}
  \caption{ \textbf{Time evolution of a localized wavepacket.} The numerically evaluated, inverse-averaged spread $\sigma(t)$ of an initially localized wavepacket for systems of size $L=400$ is shown for \fc{a} fixed $\Lambda=2J$, $\tau J=100$, \fc{b} $\Lambda=2J$, $W=2J$, and \fc{c} $W=8J$, $\tau J=100$. Dotted lines indicate diffusive expansion $\sigma(t) \sim \sqrt{t}$. Errorbars are obtained from the sample average of 250 noise and disorder realizations.
  } 
  \label{fig:spread}
\end{figure*}

\textbf{\textit{Model.---}}We consider non-interacting fermions in one dimension, subject to a static disorder potential and time dependent noise, as described by
\begin{equation}
  H = - J \sum_{\langle ij\rangle} c_i^\dag c_j^\nag + \sum_i [\epsilon_i+\xi_i(t)]c_i^\dag c_i^\nag ,
  \label{eq:h}
\end{equation}
where $J$ represents the tunneling matrix element and $c_i^\dag$ ($c_i^\nag$) creates (destroys) an electron on lattice site $i$. The on-site energies $\epsilon_i$ are uncorrelated, and are drawn from a Gaussian distribution of width $W$ and zero mean. The noise $\xi_i(t)$ is characterized by its strength $\Lambda$ and correlation time $\tau$. We consider spatially uncorrelated noise generated by an Ornstein-Uhlenbeck process~\cite{gardiner1985handbook} with temporal correlations
\begin{equation}
C(t)=\langle \xi_i(t) \xi_i(0) \rangle =  \Lambda^2 \exp[-|t|/\tau].  
\label{eq:noise}
\end{equation}
We will be interested in how transport changes as a function of the disorder strength $W$ and the noise strength $\Lambda$ as its correlation time is tuned from the Markovian, white noise limit, $\tau\to 0$, to the limit of quasistatic noise, $\tau \to \infty$. 

\textbf{\textit{Perturbative treatment.---}}
We can analytically explore the noise-induced dynamics, working in the deeply localized limit where the single-particle hopping is the smallest energy scale. (However, the ratios of the other three scales $W, \Lambda, 1/\tau$ can be arbitrary, so long as each is much larger than $J$.) In this limit, nearest-neighbor hops dominate transport; moreover, the system dephases completely between successive hops, so transport is purely incoherent. Thus, the system can be modeled as a classical one-dimensional hopping model, with rates given by the incoherent nearest-neighbor hopping rate. This rate can be computed through a treatment of a two-site problem~\cite{amir_classical_2009}.
%
One can solve the equations of motion generated by the Hamiltonian \eqw{eq:h} perturbatively in the hopping $J \ll \Lambda, W$. For zero hopping, each site simply accumulates phase, and its wavefunction amplitude at time $t$, denoted $A^0_j$, is given by $A^0_j(t) = A_j^0 e^{-i\epsilon_i t -i\phi_j(t)}$, where $\phi_j(t) = \int_0^t \xi_j(t') dt'$. To describe transport, we expand the equations of motion to the lowest nontrivial order in the hopping, resulting in the following rate equation for the probability distribution $p_j \equiv |A_j|^2$ for the particle position at time $t$, see Supplemental Material~\cite{supp}:
\begin{align}
  \frac{d p_j}{dt}\! =\! \Gamma_{j,j+1} p_{j+1}\!+\! \Gamma_{j,j-1} p_{j-1} \! - \!(\Gamma_{j+1,j}\!+\!\Gamma_{j-1,j})p_{j}
  \label{eq:diff}
\end{align}
with a locally varying rate $\Gamma_{i,j}=\Gamma(\epsilon_i-\epsilon_j)$ that depends on the \emph{energy difference} between neighboring sites $i$ and $j$: 
 $\Gamma(\omega) = 2J^2\int_0^\infty dt \cos(\omega t) \, |C^\phi(t)|^2,$
where $C^\phi(t)$ is the phase correlation function $ C^\phi(t)=\langle e^{-i\phi_j(t)}e^{i\phi_j(0)} \rangle  = e^{-\int_0^t (t-t') C(t') dt'} $ with the noise correlation function $C(t)$, \eq{eq:noise}, and we have performed a Gaussian average over noise trajectories. For our specific noise model and $\Lambda \tau \gtrsim 1$, the rate $\Gamma(\omega)$ has the form

\begin{equation}\label{corr}
\frac{\Gamma(\omega)}{2J^2} = \left\{ \begin{array}{ll} \frac{\Lambda}{\omega^2 + \Lambda^2} & \omega < \tau^{-1} \\ 
\Lambda^{-1} e^{-\omega^2/(4 \Lambda^2)} &  \tau^{-1} < \omega < 2 \Lambda \sqrt{\log(\Lambda \tau)} \\
\frac{\Lambda^2}{2 \tau \omega^4} & \omega > 2 \Lambda \sqrt{\log(\Lambda \tau)}  \end{array}   \right.
\end{equation}
Note that \eq{eq:diff} has the form of a random walk with locally varying transition rates. In the disorder-free limit~\cite{amir_classical_2009}, $\Gamma$ has no spatial dependence, and Eq.~\eqref{eq:diff} reduces to a discretized diffusion equation with a diffusion constant $\Gamma(0)$. 

\textbf{\textit{Subdiffusive regime.---}}In the disordered system, the transition rate $\Gamma_{ij}$ between a particular pair of neighboring sites depends on their energy difference $\omega$ through~\eq{corr}. For very small or very large $\omega$, the rate decreases polynomially with $\omega$. However, in the intermediate regime, which exists only for sufficiently large $\tau$, $\Gamma(\omega)$ decreases very rapidly as $\omega$ increases. This rapid decrease, as we now discuss, is the origin of anomalous diffusion.

To this end, we estimate the density of very weak links in this regime. Recall that the on-site energies are Gaussian distributed. Then the cumulative distribution function of finding a bottleneck, defined by the transition rate being smaller than a certain threshold $\Gamma_0$, follows a power-law relation~\cite{supp}
\begin{equation}
 P(\Gamma<\Gamma_0) \sim \left( \frac{\Lambda \Gamma_0}{2J^2} \right)^\frac{\Lambda^2}{W^2}.
 \label{eq:cdf}
\end{equation}
As noted above, we can directly map our problem to a classical rate equation (or resistor network) with random rates (conductances), across each nearest-neighbor link, drawn from the distribution~\eqref{eq:cdf}. For resistors that are power-law distributed $P(R)=(R_0/R)^{\mu+1}$, the mean resistance is finite for $\mu>1$ (leading to regular diffusion) but ill-defined for $\mu<1$ (leading to subdiffusion~\cite{bouchaud_anomalous_1990, agarwal_anomalous_2015}). Our rate distribution corresponds to a heavy-tailed ($\mu < 1$) resistance distribution and thus to subdiffusion when $\Lambda < W$.

\textbf{\textit{Crossover to diffusion.---}}Within our noise model, there are two mechanisms that result in a crossover to diffusion at late times. We call these respectively the ``variable-range hopping'' (VRH) and ``ultraviolet'' (UV) mechanisms. We begin by discussing the VRH mechanism, which is more generally applicable. This mechanism involves processes that avoid a bottleneck by tunneling virtually through it. Crucially, for a site to act as a bottleneck, \emph{all} transitions out of it, not just nearest-neighbor hops, must be blocked. The matrix element for an $n$-site virtual process is $J (J/W)^{n - 1}$, and the corresponding incoherent rate is given by 
%
$\Gamma_i^{(n)} \simeq \frac{2 J^2}{\Lambda} \left( \frac{J}{W} \right)^{2(n - 1)} \exp\left[-\frac{\omega^2}{4 \Lambda^2}\right].$
%
For a site to act as a bottleneck we require that $\prod_n \Gamma_i^{(n)} \alt \Gamma_0$, i.e., each link must independently act as a bottleneck. In effect, this product only runs over $n \leq n^* = \log(\Gamma_0 \Lambda / 2W^2) / 2\log(J/W)$, as more distant links are slower than $\Gamma_0$ regardless of the energy difference $\omega$~\cite{supp}. The probability of finding a series of such sites can be estimated (Supplemental Material~\cite{supp}) as 

\begin{equation}
 \tilde P(\Gamma_0|n^*) \sim \exp\left[-c \log^2 \frac{\Gamma_0 \Lambda}{2W^2} \right], 
 \label{eq:bottle}
\end{equation}
with a constant $c \simeq \Lambda^2 / (4 W^2 \log [W/J])$. This probability decays slightly faster than a powerlaw in $1/\Gamma_0$ and, hence, bottlenecks are asymptotically always sufficiently rare such that diffusion is recovered. Specifically, as the mean inverse transition rate (i.e., ``resistance'') is well-defined, we can compute the asymptotic diffusion constant by taking the inverse of this mean resistance. For $W\gg\Lambda$ we find~\cite{supp}

\begin{equation}\label{DVRH}
D_{\mathrm{VRH}} \simeq  \frac{W}{\sqrt{\pi \log[W/J]}} \left(\frac{J}{W} \right)^{W^2/\Lambda^2}.
\end{equation}

We now turn to the ``ultraviolet'' mechanism. 
Within our noise model~\eqref{eq:noise}, the correlation function~\eqref{corr} crosses over from a Gaussian to a power law, $1/\omega^4$, at large frequencies $\omega > 2\Lambda \sqrt{\log (\Lambda\tau)}$. The incoherent transition rate for pairs of sites with detuning $|\epsilon_i - \epsilon_j| > 2\Lambda \sqrt{\log (\Lambda\tau)}$ is not suppressed strongly with their detuning, and they do not bottleneck transport. The overall diffusion constant is set by the weakest \emph{common} links, for which $\omega \simeq 2\Lambda \sqrt{\log \Lambda \tau}$. The density of these links is $\exp[-\Lambda^2 \log(\Lambda \tau)/2W^2]$, and the rate across each is $2J^2/(\Lambda^2 \tau)$. Thus the effective UV diffusion constant scales as

\begin{equation}\label{DUV}
D_{\mathrm{UV}} \simeq \frac{2J^2}{\Lambda} \frac{1}{(\Lambda \tau)^{1 - (\Lambda / W)^2}}.
\end{equation}
Within our noise model, the asymptotic diffusion coefficient is set by $\text{max}( D_{\mathrm{VRH}},D_{\mathrm{UV}})$. 
However, the power-law regime in~\eqref{corr} originates from the ``cuspy'' short-time behavior of the noise correlation function~\eqw{eq:noise}. This feature is model-dependent, and indeed is absent for noise generated, e.g., by the dynamics of a finite-bandwidth quantum system. In systems where the noise correlation function is analytic at short times, diffusion is solely due to the VRH mechanism.

%
%
%
%

\textbf{\textit{Numerical results.---}}We quantitatively study the subdiffusive transport by performing exact numerical simulations of a particle localized in the center of our system. We first compute stochastic noise trajectories based on the Ornstein-Uhlenbeck process. Second, we numerically solve the equations of motion set by Hamiltonian \eqw{eq:h}. We consider systems of size $L=400$ and times to $tJ=10^4$. A typical example for the spread $\sigma(t) = \sqrt{\langle  \hat x^2 \rangle-\langle\hat x \rangle^2}$ is shown in \figc{fig:schematic}{c}. Here, the expectation values $\langle \dots \rangle$ are taken with respect to the time evolved wave function $|\psi(t)\rangle$. At times $tJ \lesssim 1$, the expansion of the wavepacket is ballistic. At later times the spread crosses over to sub-diffusive behavior $\sigma(t) \sim t^\beta$. In that regime, the \emph{direct} sample average of the spread over disorder and noise realizations $\langle \sigma(t) \rangle$, solid line, and the \emph{inverse} average of the inverse spread $1/\langle \sigma^{-1}(t) \rangle$, dashed lines, strongly disagree. This is a manifestation of the probability distribution \eqw{eq:cdf} having ill defined moments. The apparent subdiffusion exponent $\beta$ increases with time and slowly approaches the diffusive limit $\beta=1/2$ at late times $tJ \sim 10^4$. 

Simulations of the wave-packet spread $\sigma(t)$ are shown in \fig{fig:spread} for a range of parameters. Generally, we observe (i) an initial ballistic expansion, followed by (ii) an intermediate subdiffusive regime that gradually crosses over to (iii) diffusion. With increasing disorder, $\sigma(t)$ decreases and the crossover to diffusive transport is pushed to later times, \figc{fig:spread}{a}. Moreover, with increasing noise correlation time $\tau$, the intermediate subdiffusive regime is extended, leading to a decrease of the asymptotic diffusion constant with $\tau$,  \figc{fig:spread}{b}. This suggests that for the relevant parameters subdiffusion is cut off by the ``ultraviolet'' mechanism~\eqref{DUV}. Finally, at strong disorder, transport is facilitated with increasing noise strength, \figc{fig:spread}{c}. At weak disorder, however, noise \emph{impedes} transport (data not shown). These qualitative findings are fully consistent with expectations from perturbation theory.

\begin{figure}
  \centering 
  \includegraphics[width=.48\textwidth]{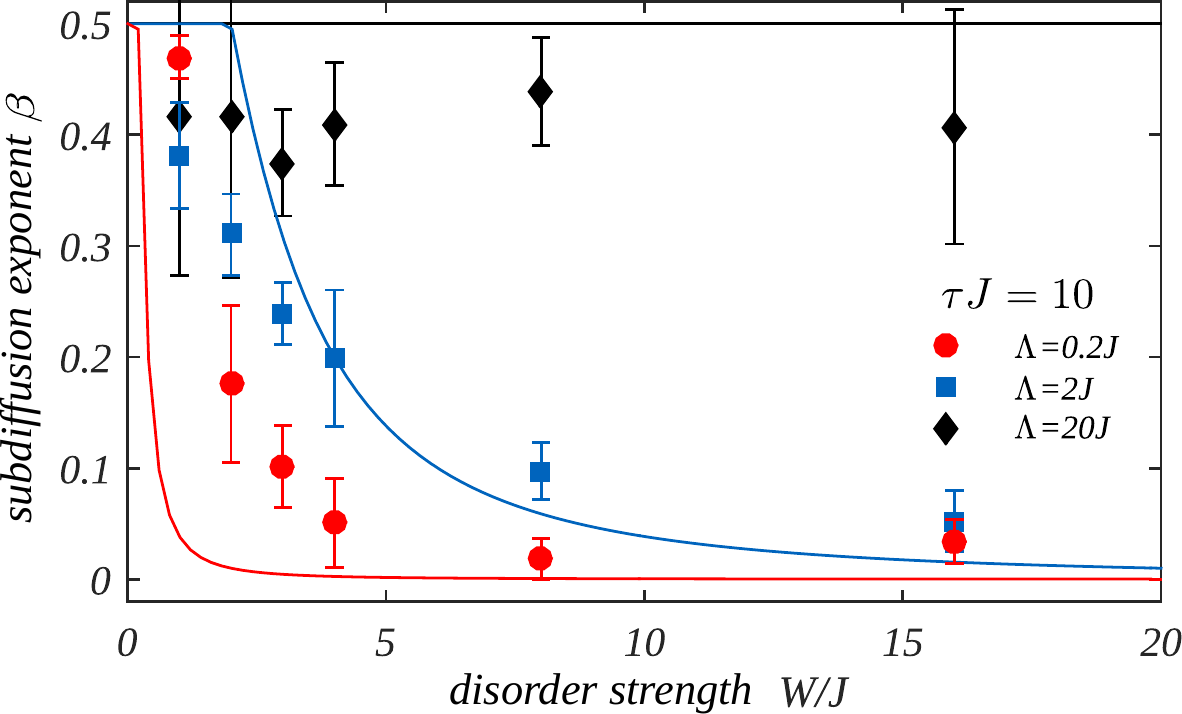}
  \caption{ \textbf{Short-time subdiffusion exponent.} The powerlaw exponent $\beta$ characterizing the initial subdiffusive transport $\sigma(t)\sim t^\beta$ is extracted from fitting the numerical data in the range $1<tJ<\tau J$ for $\tau J = 10$. In the weak noise limit $\Lambda \lesssim J$, the exponent depends strongly on the disorder strength $W$ approaching zero with increasing $W$ whereas it is constant for strong-noise $\Lambda=20J$. For weak disorder strength $W\to 0$, $\beta$ universally approaches within the errorbars the limit of diffusive transport with $\beta=1/2$. Solid lines indicate our estimate $\beta=\Lambda^2/(\Lambda^2+W^2)$.
  } 
  \label{fig:subexp}
\end{figure}

We evaluate the subdiffusion exponent $\sigma(t) \sim t^\beta$ by fitting the numerical data in the regime $1<tJ<\tau J$, \fig{fig:subexp}. The small range is chosen to capture the exponent at the onset of the subdiffusive regime. For weak noise and strong disorder, the subdiffusion exponent $\beta$ is near zero. When lowering the disorder strength, $\beta$ strongly increases and approaches the diffusive limit $\beta \to 1/2$. By contrast at large disorder, $\beta$ sets off at a larger value and quickly saturates. We obtain an estimate for $\beta$ by relating it to the exponent of the cumulative distribution function \eqw{eq:cdf} as $\beta={\Lambda^2}/({\Lambda^2+W^2})$~\cite{agarwal_anomalous_2015}; indicated by solid lines in \fig{fig:subexp}.
The data qualitatively reproduces the predicted trend but slight quantitative differences are present. Such discrepancies are not unexpected as our theoretical analysis is valid to lowest order in $J/W, J/\Lambda$, and these parameters are not small in the numerically accessible regime. 

\begin{figure}
  \centering 
  \includegraphics[width=.48\textwidth]{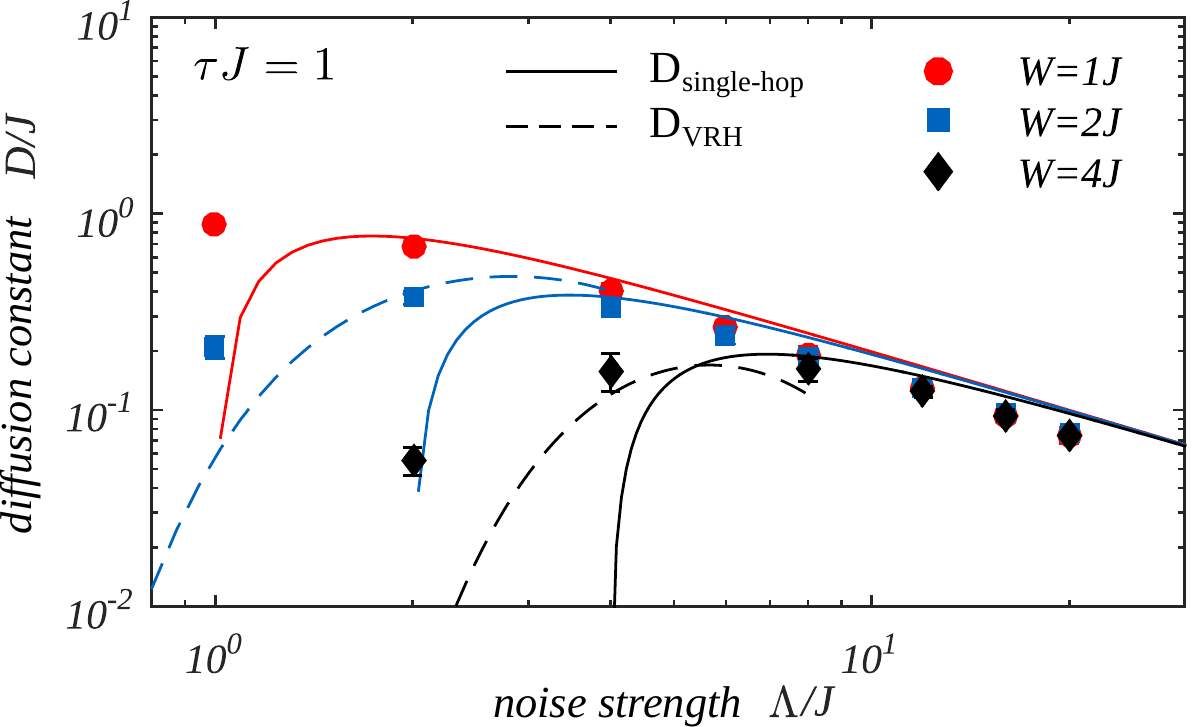}
  \caption{ \textbf{Asymptotic diffusion constant.} The diffusion constant $D$ evaluated from the asymptotic spread of the wavepacket $\sigma(t\to\infty)=\sqrt{2 D t}$ is shown as a function of the the noise strength $\Lambda$ for different values of the disorder strength $W$ and fixed noise correlation time $\tau J=1$, symbols. The numerical data is compared to the single-hop model valid for $\Lambda \gtrsim W$, solid lines, and the variable-range hopping model of \eq{DVRH} valid for $\Lambda \lesssim W$, dashed lines. 
  } 
  \label{fig:diff}
\end{figure}

From the long-time asymptotics of the spread $\sigma(t)$, we extract the diffusion constant $\sigma(t\to\infty)=\sqrt{2Dt}$ for different values of the noise and disorder strength  at fixed noise correlation time $\tau J=1$ (\fig{fig:diff}). For strong noise compared with disorder, the diffusion constant is largely disorder-independent, and \emph{decreases} with increasing noise as $\sim 1/\Lambda$, \eq{corr}, consistent with Ref.~\cite{amir_classical_2009,znidaric_transport_2013}. In the strong noise limit, $\Lambda \gg W$, diffusion is induced already by nearest neighbor hops, leading to $D_\text{single-hop} \sim 2J^2(1 -  W^2/\Lambda^2)/\Lambda$ (solid lines)~\cite{supp}.
In the subdiffusive regime, $\Lambda<W$, it is challenging to propagate to sufficiently long times to see the eventual crossover to diffusion. However, we were able to extract a few data points in that limit, and observe a reversed dependence: noise assists diffusion rather than impeding it as predicted by the variable range processes, \eq{DVRH}, (dashed lines).

\textbf{\textit{Discussion}}.---How robust are our conclusions to adding interactions, and to more general forms of correlated noise? Adapting our results to interacting, many-body localized systems coupled to noise is straightforward in principle. Qualitatively, the main difference is that there are many more ways for an interacting system to ``escape'' a bottleneck: in addition to longer-range hops, the system can undergo many-particle rearrangements, which have a larger phase space~\cite{Gopalakrishnan15}. Thus the variable-range hopping mechanism will be more effective, giving rise to a smaller subdiffusive window and a larger asymptotic diffusion constant. In the Supplemental Material support these expectations by studying transport in \emph{interacting} and \emph{localized} systems using numerical simulations based on Lanczos exact diagonalization~\cite{supp}.  

Interacting localized systems coupled to Markovian baths have been shown to exhibit a stretched-exponential decay of the contrast of an initial density-wave pattern~\cite{fischer2016dynamics, levi2016robustness, medvedyeva2016influence}. We find similar behavior in our system with a stretching exponent that is independent of the noise correlation time $\tau$ for weak noise $\Lambda \lesssim J$ but depends strongly on $\tau$ for large noise $\Lambda \gtrsim J$~\cite{supp}. 

Our perturbative analysis suggests that our numerical results should be sensitive to the short-time correlations of the noise, which are nonuniversal. In particular, noise emanating from a physical system with a finite bandwidth will decay as a Gaussian, rather than an exponential, on timescales that are short compared with the bandwidth. Thus, the ``ultraviolet'' mechanism should be absent in such systems. This interdependence of slow and fast processes has also been observed in a mean-field treatment of the MBL transition~\cite{gopalakrishnan2014mean}, and appears to be a generic phenomenon, reminiscent of ``UV-IR mixing'' in field theory~\cite{minwalla2000noncommutative}. Extending our numerical studies to more general forms of colored noise is an important direction, which we shall explore in future work. 

\textbf{\textit{Outlook}}.---We have studied noise-induced transport in disordered quantum systems. We have argued that for slowly fluctuating noise, transport is governed by an incoherent hopping model with an \emph{emergent} broad distribution of hopping rates, causing anomalous diffusion on intermediate timescales, and regular diffusion (with a strongly suppressed diffusion constant) at late times. The subdiffusive regime is parametrically large when the noise correlation time $\tau$ is long, so that $\Lambda \ll W \ll \Lambda \sqrt{\log(\Lambda \tau)}$. As this regime grows slowly with $\tau$, numerical simulations cannot access the regime where perturbation theory is quantitatively reliable. Nevertheless, simulations clearly show the predicted trends, specifically subdiffusive intermediate-time dynamics and a late-time crossover to diffusion.

Our approach paves the way for developing a self-consistent theory for the thermal phase in disordered interacting quantum systems where interactions can be treated by a self-consistent Hartree-Fock decoupling. It has been argued that, within a Hartree-Fock treatment, interacting and disordered bosons exhibit subdiffusive transport~\cite{lucioni_observation_2011, shepelyansky_delocalization_1993, flach_universal_2009}. Hence, it would be interesting to classify the effective noise spectrum in such mean-field bosonic systems and analyze the anomalous transport using the framework presented here. Furthermore, such a self-consistent theory can also be developed with the prospect of studying the response of a fully many-body localized system coupled to a bath. Having technical approaches at hand, which go beyond conventional exact diagonalization of small quantum systems, will help to provide further insight in the many-body localized phase and its breakdown.

\textit{\textbf{Acknowledgments.---}}We thank K. Agarwal, A. Amir, S. Choi, E. Demler, M. Lukin, V. Oganesyan, and W. Witt for many useful discussions. We acknowledge support from Technical University of Munich - Institute for Advanced Study, funded by the German Excellence Initiative and the European Union FP7 under grant agreement 291763, from the DFG grant No. KN 1254/1-1, and the Walter Burke Foundation at Caltech.

\newpage
\clearpage
\appendix

\setcounter{figure}{0}
\setcounter{equation}{0}

\renewcommand{\thepage}{S\arabic{page}} 
\renewcommand{\thesection}{S\arabic{section}} 
\renewcommand{\thetable}{S\arabic{table}}  
\renewcommand{\thefigure}{S\arabic{figure}} 
\renewcommand{\theequation}{S\arabic{equation}} 

\onecolumngrid

\section{\large{Supplemental Material: \\Noise-induced subdiffusion in strongly localized quantum systems}}

\section{Perturbative treatment}

We discuss how to establish analytical insights in the noise-induced dynamics by a perturbative treatment in small hopping $J \ll \Lambda, W$. The equations of motion for the annihilation operator $c_j$ set by Hamiltonian \eqs{1} read
\begin{equation}
i \frac{d c_j}{dt} = -J (c_{j-1}+c_{j+1}) + [\epsilon_j +\xi_j(t)]c_j.
\end{equation}
We solve these equations order by order in the hopping $J$~\cite{amir_classical_2009}. In the absence of interactions we can represent the quantum operator $c_j$ by a complex amplitude $A_j$. The dynamics of the wave function amplitude to leading order $A_j^0$ is determined by
\begin{equation}
i\frac{d A_j^0}{dt} = [\epsilon_j +\xi_j(t)]A_j^0,
\end{equation}
which describes the accumulation of phase
\begin{equation}
A_j^0(t) = A_j^0 e^{-i\epsilon_j t - i \int_0^t \xi_j(t') dt'} = A_j^0 e^{-i\epsilon_j t} e^{-i\phi_j(t)}.
\end{equation}
To leading order transport is absent. However, it is restored by evaluating the next-to-leading order correction 
\begin{equation}
i \frac{dA_i^1}{dt} - [\epsilon_i+\xi_i(t)]A_i^1 = -J (A^0_{j+1}+A^0_{j-1}) .
\end{equation}
Introducing $\mu_j = e^{i \phi_j(t)}$, we rewrite the equation as $\frac{i}{\mu_j}\frac{d(A_i^1 \mu_j)}{dt}= -J (A^0_{j+1}+A^0_{j-1})$, which has the solution
\begin{equation}
A_j^1(t) = A_j^0(t) + \frac{iJ}{\mu_j(t)} \int_0^t dt' \mu_j(t') [A_{j+1}^0(t')+A_{j-1}^0(t')].
\end{equation}
Next, we express the Heisenberg equations of motion in terms of the probability distribution $p_j = |A_j|^2$
\begin{equation}
\frac{d p_j}{dt} = - 2J \im [A_j^* A_{j+1}+A_j^* A_{j-1}].
\end{equation}
Plugging in the next-to-leading order result for the amplitudes $A_j^1$ and taking the average over the noise, we obtain the rate equation {(3)} for the probability distribution  with the rates
\begin{equation}
\Gamma(\epsilon_i-\epsilon_j) = 2J^2 \re \langle \int_0^t dt' e^{-i[\phi_{j}(t)-i\phi_{j}(t')]}e^{i(\phi_i(t)-i\phi_i(t'))} \rangle =  2J^2  \int_0^t dt' \cos[(\epsilon_j-\epsilon_i)t'] \, |C^\phi(t')|^2 .
\end{equation}
Hence, in the asymptotic limit, $t\to\infty$, the rate is determined by the Fourier transform of the kernel $|C^\phi(t)|^2 = \exp \left[ -2 \int_0^t (t-x) C(x) dx \right]$ evaluated at the energy difference of the neighboring sites. We evaluate the rate $\Gamma(\omega)$ for our noise model, \eqs{2}, which in the strong noise limit $\Lambda \tau \gtrsim 1$ yields \eqs{5}. The rate thus exhibits an intermediate Gaussian regime that exists for large noise correlation times $\tau$. This strong decay of the rate with frequency $\omega$ leads to bottlenecks and is the origin of the subdiffusive transport.

\section{Subdiffusive transport}

The strong decay of the rate $\Gamma(\omega)$ in the intermediate Gaussian regime leads to bottlenecks. We introduce a cutoff $\Gamma_0$ and define that rates that are smaller than $\Gamma_0$ realize bottlenecks and block transport
\begin{equation}
\Gamma(\omega) = \frac{2J^2}{\Lambda} e^{-\omega^2/4\Lambda^2} < \Gamma_0 .
\end{equation}
Inverting this equation, we obtain a bound on the energy $|\omega|> 2\Lambda \sqrt{-\log \frac{\Lambda \Gamma_0}{2J^2}} \equiv2\Lambda \sqrt{-\log \tilde{\Gamma}_0} $. We first consider that diffusion is initiated by resonant processes between nearest neighbor sites. Thus the frequency $\omega$ needs to be resonant with a random variable $x$ drawn from the distribution of the nearest neighbor energy differences, which is a Gaussian of width $\sqrt{2}W$, where $W$ is the local disorder strength: $N(x,\sqrt{2}W) = \frac{1}{\sqrt{\pi}2W} e^{-x^2/4W^2}$. The cumulative probability distribution of finding rates that are smaller than the cutoff is thus
\begin{equation}
P(\Gamma<{\Gamma}_0) = P(x>2\Lambda \sqrt{-\log \tilde{\Gamma}_0}) = \int_{2\Lambda \sqrt{-\log \tilde{\Gamma}_0}}^\infty N(x,\sqrt{2}W) dx = \frac{1}{2} \erfc [\frac{\Lambda}{W} \sqrt{-\log \tilde{\Gamma}_0}].
\end{equation}
In the asymptotic limit of small $\tilde{\Gamma}_0$ we approximate $\erfc[z] \sim \frac{\exp[-z^2]}{z \sqrt{\pi}}$ and hence find that the cumulative distribution function obeys (up to logarithmic corrections) a powerlaw
\begin{equation}
P(\Gamma<{\Gamma}_0) \sim e^{\frac{\Lambda^2}{W^2} \log{\tilde{\Gamma}_0}} \sim \tilde{\Gamma}_0^{\frac{\Lambda^2}{W^2}}.
\label{eq:cdfRate}
\end{equation}
Interpreting the local transition rates as inverse resistors, we make an analogy with a random resistor network model and find subdiffusive transport when the exponent of $P(\Gamma<{\Gamma}_0)$ is less than one~\cite{agarwal_anomalous_2015,bouchaud_anomalous_1990}
\begin{equation}
\Lambda <  W .
\end{equation}

In summary, we expect subdiffusion for $ \Lambda < W < 2 \Lambda \sqrt{\log \Lambda \tau}$. Thus, $\tau$ has to be large enough to enable this anomalous transport regime. 

\section{Crossover to Diffusion}

Thus far we only considered hopping processes between nearest neighbors. However, once we find a small nearest-neighbor rate, it does not automatically mean that we do have global subdiffusion. Analogously to variable range hopping, we consider higher-order hopping processes to more distant neighbors which scale as $J (J/W)^{(n-1)}$. Only if none of these transition rates is large, the site can act as a bottleneck. Using the renormalized hopping, the transition rate at order $n$ is given by $\Gamma_i^{(n)} \simeq \frac{2 J^2}{\Lambda} \left( \frac{J}{W} \right)^{2(n - 1)} \exp\left[-\frac{\omega^2}{4 \Lambda^2}\right]$. The corresponding cumulative distribution function reads
\begin{equation}
P(\Gamma_i^{(n)}<{\Gamma}_0) = \left[ \tilde \Gamma_0  ({W}/{J})^{2(n-1)}\right]^\frac{\Lambda^2}{W^2}.
\end{equation}
The probability of finding a series of such slow sites (taking them as independent processes) is
\begin{equation}
\tilde P(\Gamma_0|n^*) = \prod_{n=1}^{n^*}\left[  \frac{\Gamma_0\Lambda}{2J^2} ({W}/{J})^{2(n-1)}\right]^\frac{\Lambda^2}{W^2},
\end{equation}
where $n^*$ characterizes the distance beyond which all rates are small compared to $\Gamma_0$ by definition. We estimate this maximum distance by
\begin{equation}
\Gamma_0= \frac{2J^2}{\Lambda}  (J/W)^{2(n^*-1)} .
\end{equation}
Solving for $n^*$ we obtain $n^* = \log(\Gamma_0 \Lambda / 2W^2) / 2\log(J/W)$. Taking this maximal distance, the probability of finding a series of slow sites is
\begin{equation}\label{distro}
\tilde P(\Gamma_0|n^*) \simeq \left( \frac{\Lambda \Gamma_0}{2W^2}\right)^{-\Lambda^2/(2W^2)} \exp\left[-\frac{\Lambda^2}{4 W^2 \log W/J} \log^2 \frac{\Gamma_0 \Lambda}{2W^2} \right],
\end{equation}
which is decaying slightly faster than a powerlaw with $1/\Gamma_0$. Therefore, bottlenecks become ineffective at asymptotically late times and subdiffusive transport crosses over to diffusion. 

\begin{figure}
	\centering 
	\includegraphics[width=.48\textwidth]{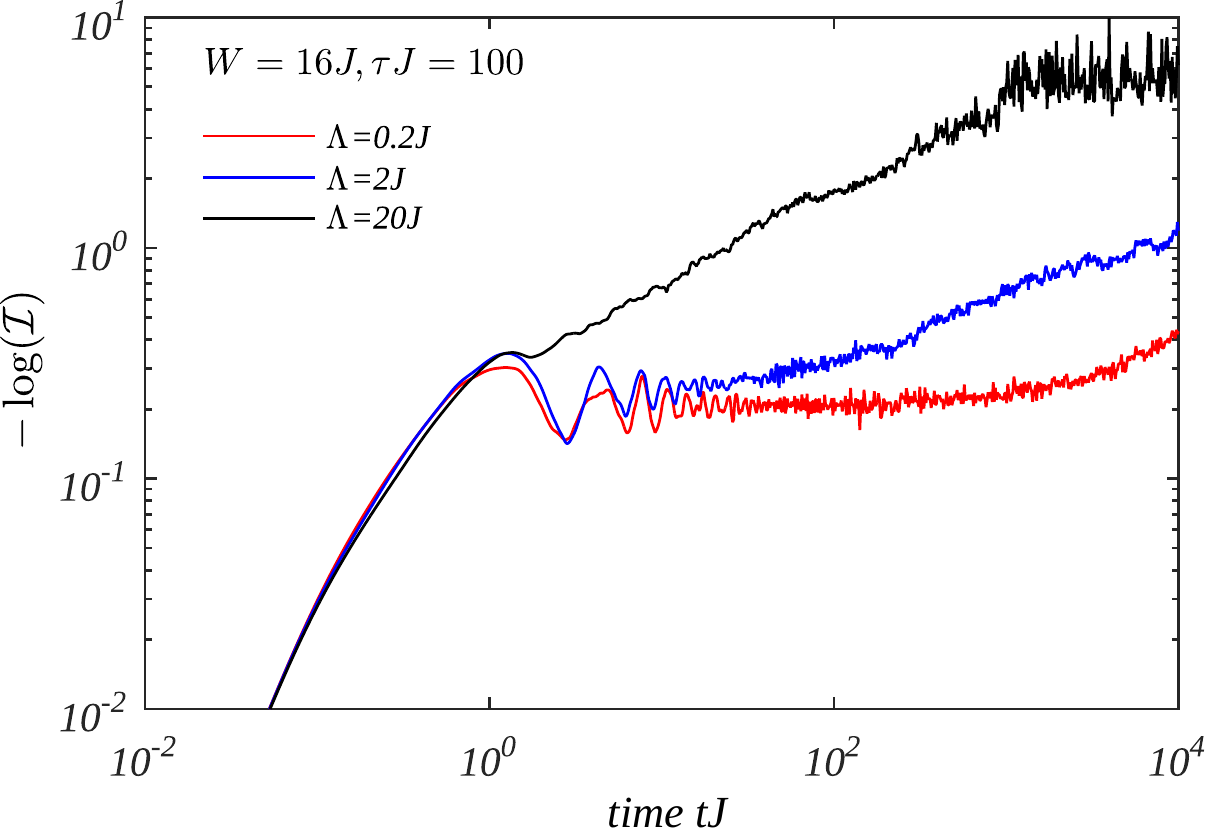}
	\caption{ \textbf{Stretched-exponential decay of the imbalance in a noisy environment.} The contrast of an initial density-wave pattern of occupied even and unoccupied odd lattice sites, denoted as imbalance $\mathcal{I}$, is shown for strong disorder $W=16J$, large noise correlation times $\tau J=100$ and three different values of the noise strength $\Lambda$. The asymptotic stretched exponential decay of the imbalance, \eq{eq:stretch}, can be inferred from plotting $-\log{\mathcal{I}}$ on a double logarithmic plot, in which the stretching exponent $\alpha$ can be directly read off from the slope of the linear growth at late times. 
	} 
	\label{fig:imbalance}
\end{figure}

We now estimate the diffusion constant, by computing the mean resistance and inverting it: via the Einstein relation, we can identify the dc conductance with the diffusion constant. Using the cumulative distribution function~\eqref{distro} for sites with decay rates smaller than $\Gamma_0$, we proceed as follows. First, we note that the ``resistance'' $R$ is identified with the inverse rate. Second, from \eq{distro}, we compute the probability density by computing the derivative of $\tilde P(\Gamma_0|n^*)$
\begin{equation}
p(R) = \frac{1}{R}  \frac{\Lambda^2}{2 W^2 \log(W/J)} \left[\log\left(\frac{2W^2 R}{\Lambda}\right)-\log\left(\frac{W}{J}\right)\right]   \tilde{P}(1/R|n^*).
\end{equation}
Using this distribution, we can estimate the mean resistance, which is given by
\begin{equation}
\langle R \rangle \simeq \frac{\sqrt{\pi \log[W/J]}}{W}  \left(\frac{W}{J}\right)^{(W/\Lambda+\Lambda/2W)^2} 
\end{equation}
from which it follows that the asymptotic diffusion coefficient is given (for large $W/\Lambda$) by

\begin{equation}
D_{\mathrm{VRH}} \sim \frac{W}{\sqrt{\pi \log[W/J]}} \left( \frac{J}{W} \right)^{W^2/\Lambda^2}.
\end{equation}

This expression only applies when $W > \Lambda$, and is only \emph{controlled} when $W \gg \Lambda$. In the opposite limit, $W \ll \Lambda$, one gets diffusion even from incoherent single-site hopping. The diffusion constant in that regime can be found by computing the average resistance due to lowest-order hops, \eq{eq:cdfRate}, which leads to the result

\begin{equation}
D_\text{single-hop} \sim \frac{2J^2}{\Lambda} (1 - W^2 / \Lambda^2),
\end{equation}
i.e., it vanishes as $\Lambda \rightarrow  W$, and then crosses over to the VRH form above.

\section{Imbalance}

\begin{figure*}
	\centering 
	\includegraphics[width=.98\textwidth]{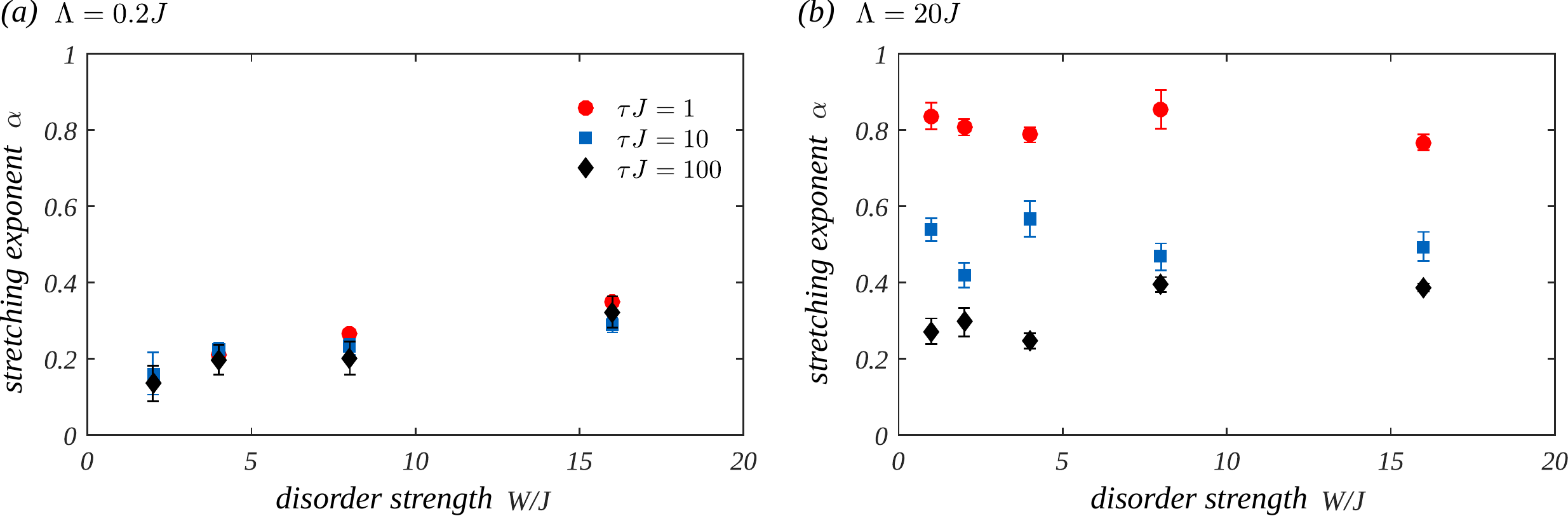}
	\caption{ \textbf{Stretching exponent of the imbalance.} The stretching exponent $\alpha$ is shown \fc{a} in the weak noise limit $\Lambda =0.2 J$ and \fc{b} in the strong noise limit $\Lambda=20J$. For weak noise, the exponent does not depend on the noise correlation time $\tau$ but depends weakly on the disorder strength $W$. By contrast, for strong noise, the stretching exponent is very sensitive to the noise correlation time $\tau$. For short correlation time $\tau J=1$ the stretching exponent is close to one, indicating a nearly exponential decay of the imbalance $\mathcal{I}$. 
	} 
	\label{fig:stretchingExp}
\end{figure*}

Many-body localized systems coupled to a Markovian bath have been shown to exhibit a large distribution of relaxation rates, which manifests itself in an asymptotic stretched exponential decay of the imbalance $\mathcal{I}$ of an initial charge density wave pattern of occupied even and unoccupied odd sites~\cite{fischer2016dynamics, levi2016robustness, medvedyeva2016influence}
\begin{equation}
\mathcal{I}(t\to \infty) = \exp\left[-({t}/{\tau})^\alpha\right],
\label{eq:stretch}
\end{equation}
where $\alpha$ is the stretching exponent. This quantity has been thoroughly investigated theoretically, since it has been used in experiments to establish the many-body localized phase~\cite{Schreiber15,bordia_periodically_2016}. Here, we show that also for non-interacting systems in a noisy environment the imbalance decays as a stretched exponential, \fig{fig:imbalance}, which is best demonstrated by plotting $-\log \mathcal{I}$ on double logarithmic scales. In such a plot the stretching exponent $\alpha$ can directly be read off from the slope of the linear curve at late times. In the weak noise limit $\Lambda=0.2J$ the imbalance remains constant up to late times $tJ\sim10^3$ and then crosses over to a stretched-exponential decay. By contrast, in the strong noise limit $\Lambda=20J$, the intermediate time plateau ceases to exist and after an initial decay on the single-particle timescale, the imbalance immediately turns to a stretched exponential. In the strong noise limit $\Lambda =20\tau$, the curve saturates at late times which we attribute to the fact that the data hits the sample noise floor, as in this regime the imbalance is already $\mathcal{I}\lesssim 10^{-4}$.

We extract the stretching exponent $\alpha$ for a broad range of parameters, \fig{fig:stretchingExp}, and find that $\alpha$ is insensitive to the noise correlation time $\tau$ in the weak noise limit $\Lambda=0.2J$ \fc{a} but depends strongly on the noise correlation time for strong noise $\Lambda = 20J$ \fc{b}. In the latter regime the stretching exponent $\alpha$ approaches values near one for fast noise $\tau J=1$, indicating an almost exponential decay, whereas for slow noise $\tau J=100$, it remains appreciably smaller than one. Such a dependence of the stretching exponent on the noise correlation time cannot be studied in a Lindblad formalism~\cite{fischer2016dynamics, levi2016robustness, medvedyeva2016influence}, which assumes a Markovian bath with vanishing noise correlation times $\tau\to0$.

\section{Noise-induced dynamics in the many-body localized phase}

To study the noise-induced dynamics in the many-body localized phase, we consider disordered and interacting electrons
\begin{equation}
H = - \frac{J}{2} \sum_{i} (c_i^\dag c_{i+1}^\nag + \text{h.c.}) + U \sum_{i} \hat n_i \hat n_{i+1} + \sum_i [\epsilon_i+\xi_i(t)]\hat n_i .
\label{eq:hInt}
\end{equation}
Except for the second term, which describes the electron-electron interactions of strength $U$ and a trivial rescaling by a factor $1/2$, the Hamiltonian is identical to \eqs{1} in the main text. We solve the quantum dynamics using Lanczos time evolution and update the Hamiltonian at each time step with a new spatial noise profile, sampled from an Ornstein-Uhlenbeck process. The initial state $\ket{\tilde \psi}$ is a random product state drawn the Haar measure. We polarize the initial state by applying the operator $\mathcal{P}= (\ket{1} \bra{1})_{L/2}$ to the random state $\ket{\psi} = \mathcal{P}\ket{\tilde \psi}$. Starting with $\ket{\psi}$ we compute the time evolution of the system using Lanczos algorithm $\ket{\psi(t)} = \mathcal{T} \exp[-i\int_0^t H(t') dt'] \ket{\psi}$ and measure the polarization decay in the center of the system: $\delta n(t) = \langle \psi(t) | \hat n_{L/2} | \psi(t) \rangle - n_0$, where $n_0$ is the static expectation value of the density that is determined by the total particle number which is conserved in our model. 

\begin{figure*}
	\centering 
	\includegraphics[width=.98\textwidth]{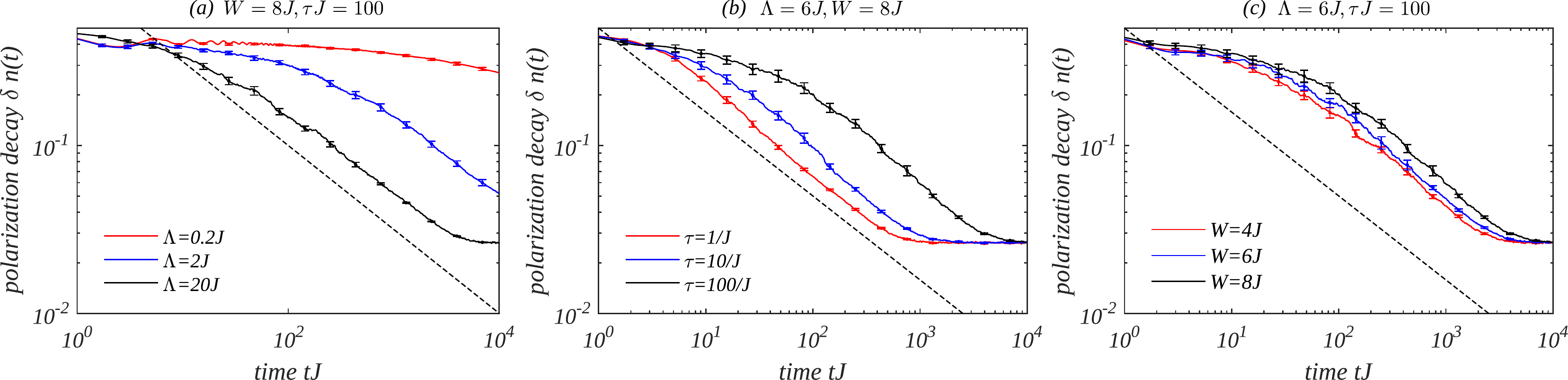}
	\caption{ \textbf{Noise-induced polarization decay for interacting and disordered fermions.} We numerically simulate the polarization decay $\delta n(t)$ for systems of size $L=19$, $N=10$ particles, interactions $U=-J$ for a broad range of disorder strength $W$, noise strength $\Lambda$, and noise correlation times $\tau$, see legends. At late times and for strong enough noise the polarization decay crosses over to diffusion $\delta n(t) \sim 1/\sqrt{t}$ (dashed black line).  
	} 
	\label{fig:szInt}
\end{figure*}

Results of the polarization decay $\Delta n(t)$ for systems of size $L=19$ with $N=10$ particles to times $tJ=10^4$ and interaction strength $U=-J$ are shown in \fig{fig:szInt} for a broad range of parameters. In order to minimize finite size effects, we consider systems with an odd number of sites. Otherwise, there would be a dangling particle in the surrounding of the initially polarized site, which leads to a late-time saturation plateau that deviates from the respective filling by a correction $\sim 1/L$.  

In the absence of noise, the system is in the many-body localized phase for the chosen parameters. However, our data shows that noise inevitably induces delocalization. For strong noise $\Lambda=20J$, \figc{fig:szInt}{a}, we find that the system quickly approaches diffusive dynamics, as described by a $1/\sqrt{t}$ decay of $\delta n(t)$. At very late times, $tJ\sim 10^4$, the response saturates to a finite value which is a consequence of the finite system size. For weaker noise, $\Lambda = 2J$, it takes the system longer to approach the diffusive regime, and for extremely weak noise $\Lambda = 0.2 J$ the polarization has almost not decayed on the simulated time scales. These numerical findings, are in agreement with our expectations discussed in the main text. In contrast to the non-interacting system, the crossover from subdiffusion to diffusion occurs more gradually, resulting from the many possible decay channels enabled by multi-particle rearrangements that are not allowed in the absence of interactions. Yet, at late times, diffusion sets the dynamics. 

The polarization decay $\delta n(t)$, is shown in \figc{fig:szInt}{b} for fixed noise strength $\Lambda=6J$ and disorder $W=8J$, for different noise correlation time $\tau$. After some initial dynamics, the polarization decay approaches the diffusive $1/\sqrt{t}$ limit. Finally, for fixed noise strength $\Lambda=6J$, and noise correlation time $\tau J=100$, \fc{c}, the system crosses over to diffusion irrespective of the disorder strength.

\begin{thebibliography}{48}%
	\makeatletter
	\providecommand \@ifxundefined [1]{%
		\@ifx{#1\undefined}
	}%
	\providecommand \@ifnum [1]{%
		\ifnum #1\expandafter \@firstoftwo
		\else \expandafter \@secondoftwo
		\fi
	}%
	\providecommand \@ifx [1]{%
		\ifx #1\expandafter \@firstoftwo
		\else \expandafter \@secondoftwo
		\fi
	}%
	\providecommand \natexlab [1]{#1}%
	\providecommand \enquote  [1]{``#1''}%
	\providecommand \bibnamefont  [1]{#1}%
	\providecommand \bibfnamefont [1]{#1}%
	\providecommand \citenamefont [1]{#1}%
	\providecommand \href@noop [0]{\@secondoftwo}%
	\providecommand \href [0]{\begingroup \@sanitize@url \@href}%
	\providecommand \@href[1]{\@@startlink{#1}\@@href}%
	\providecommand \@@href[1]{\endgroup#1\@@endlink}%
	\providecommand \@sanitize@url [0]{\catcode `\\12\catcode `\$12\catcode
		`\&12\catcode `\#12\catcode `\^12\catcode `\_12\catcode `\%12\relax}%
	\providecommand \@@startlink[1]{}%
	\providecommand \@@endlink[0]{}%
	\providecommand \url  [0]{\begingroup\@sanitize@url \@url }%
	\providecommand \@url [1]{\endgroup\@href {#1}{\urlprefix }}%
	\providecommand \urlprefix  [0]{URL }%
	\providecommand \Eprint [0]{\href }%
	\providecommand \doibase [0]{http://dx.doi.org/}%
	\providecommand \selectlanguage [0]{\@gobble}%
	\providecommand \bibinfo  [0]{\@secondoftwo}%
	\providecommand \bibfield  [0]{\@secondoftwo}%
	\providecommand \translation [1]{[#1]}%
	\providecommand \BibitemOpen [0]{}%
	\providecommand \bibitemStop [0]{}%
	\providecommand \bibitemNoStop [0]{.\EOS\space}%
	\providecommand \EOS [0]{\spacefactor3000\relax}%
	\providecommand \BibitemShut  [1]{\csname bibitem#1\endcsname}%
	\let\auto@bib@innerbib\@empty
	\bibitem [{\citenamefont {Lee}\ and\ \citenamefont
		{Ramakrishnan}(1985)}]{lee_RMP}%
	\BibitemOpen
	\bibfield  {author} {\bibinfo {author} {\bibfnamefont {Patrick~A}\
			\bibnamefont {Lee}}\ and\ \bibinfo {author} {\bibfnamefont {TV}~\bibnamefont
			{Ramakrishnan}},\ }\bibfield  {title} {\enquote {\bibinfo {title} {Disordered
				electronic systems},}\ }\href {\doibase 10.1103/RevModPhys.57.287} {\bibfield
		{journal} {\bibinfo  {journal} {Rev. Mod. Phys.}\ }\textbf {\bibinfo
			{volume} {57}},\ \bibinfo {pages} {287} (\bibinfo {year} {1985})}\BibitemShut
	{NoStop}%
	\bibitem [{\citenamefont {Kramer}\ and\ \citenamefont
		{MacKinnon}(1993)}]{kramer_RMP}%
	\BibitemOpen
	\bibfield  {author} {\bibinfo {author} {\bibfnamefont {Bernhard}\
			\bibnamefont {Kramer}}\ and\ \bibinfo {author} {\bibfnamefont {Angus}\
			\bibnamefont {MacKinnon}},\ }\bibfield  {title} {\enquote {\bibinfo {title}
			{Localization: theory and experiment},}\ }\href {\doibase
		10.1088/0034-4885/56/12/001} {\bibfield  {journal} {\bibinfo  {journal} {Rep.
				Prog. Phys.}\ }\textbf {\bibinfo {volume} {56}},\ \bibinfo {pages} {1469}
		(\bibinfo {year} {1993})}\BibitemShut {NoStop}%
	\bibitem [{\citenamefont {Anderson}(1958)}]{anderson_absence_1958}%
	\BibitemOpen
	\bibfield  {author} {\bibinfo {author} {\bibfnamefont {P.~W.}\ \bibnamefont
			{Anderson}},\ }\bibfield  {title} {\enquote {\bibinfo {title} {Absence of
				diffusion in certain random lattices},}\ }\href {\doibase
		10.1103/PhysRev.109.1492} {\bibfield  {journal} {\bibinfo  {journal} {Phys.
				Rev.}\ }\textbf {\bibinfo {volume} {109}},\ \bibinfo {pages} {1492--1505}
		(\bibinfo {year} {1958})}\BibitemShut {NoStop}%
	\bibitem [{\citenamefont {Fleishman}\ and\ \citenamefont
		{Anderson}(1980)}]{fleishman}%
	\BibitemOpen
	\bibfield  {author} {\bibinfo {author} {\bibfnamefont {L.}~\bibnamefont
			{Fleishman}}\ and\ \bibinfo {author} {\bibfnamefont {P.~W.}\ \bibnamefont
			{Anderson}},\ }\bibfield  {title} {\enquote {\bibinfo {title} {Interactions
				and the anderson transition},}\ }\href {\doibase 10.1103/PhysRevB.21.2366}
	{\bibfield  {journal} {\bibinfo  {journal} {Phys. Rev. B}\ }\textbf {\bibinfo
			{volume} {21}},\ \bibinfo {pages} {2366--2377} (\bibinfo {year}
		{1980})}\BibitemShut {NoStop}%
	\bibitem [{\citenamefont {Basko}\ \emph {et~al.}(2006)\citenamefont {Basko},
		\citenamefont {Aleiner},\ and\ \citenamefont
		{Altshuler}}]{basko_metalinsulator_2006}%
	\BibitemOpen
	\bibfield  {author} {\bibinfo {author} {\bibfnamefont {{D.M.}}\ \bibnamefont
			{Basko}}, \bibinfo {author} {\bibfnamefont {{I.L.}}\ \bibnamefont {Aleiner}},
		\ and\ \bibinfo {author} {\bibfnamefont {{B.L.}}\ \bibnamefont {Altshuler}},\
	}\bibfield  {title} {\enquote {\bibinfo {title} {Metal–insulator transition
			in a weakly interacting many-electron system with localized single-particle
			states},}\ }\href {\doibase 10.1016/j.aop.2005.11.014} {\bibfield  {journal}
	{\bibinfo  {journal} {Ann. Phys. (N.Y.)}\ }\textbf {\bibinfo {volume}
		{321}},\ \bibinfo {pages} {1126--1205} (\bibinfo {year} {2006})}\BibitemShut
{NoStop}%
\bibitem [{\citenamefont {Gornyi}\ \emph {et~al.}(2005)\citenamefont {Gornyi},
	\citenamefont {Mirlin},\ and\ \citenamefont
	{Polyakov}}]{gornyi_interacting_2005}%
\BibitemOpen
\bibfield  {author} {\bibinfo {author} {\bibfnamefont {I.~V.}\ \bibnamefont
		{Gornyi}}, \bibinfo {author} {\bibfnamefont {A.~D.}\ \bibnamefont {Mirlin}},
	\ and\ \bibinfo {author} {\bibfnamefont {D.~G.}\ \bibnamefont {Polyakov}},\
}\bibfield  {title} {\enquote {\bibinfo {title} {Interacting electrons in
		disordered wires: Anderson localization and low-t transport},}\ }\href
{\doibase 10.1103/PhysRevLett.95.206603} {\bibfield  {journal} {\bibinfo
		{journal} {Phys. Rev. Lett.}\ }\textbf {\bibinfo {volume} {95}},\ \bibinfo
	{pages} {206603} (\bibinfo {year} {2005})}\BibitemShut {NoStop}%
\bibitem [{\citenamefont {Nandkishore}\ and\ \citenamefont
	{Huse}(2015)}]{nandkishore_mbl_2015}%
\BibitemOpen
\bibfield  {author} {\bibinfo {author} {\bibfnamefont {Rahul}\ \bibnamefont
		{Nandkishore}}\ and\ \bibinfo {author} {\bibfnamefont {David~A.}\
		\bibnamefont {Huse}},\ }\bibfield  {title} {\enquote {\bibinfo {title}
		{Many-body localization and thermalization in quantum statistical
			mechanics},}\ }\href {\doibase 10.1146/annurev-conmatphys-031214-014726}
{\bibfield  {journal} {\bibinfo  {journal} {Annu. Rev. Condens. Matter}\
	}\textbf {\bibinfo {volume} {6}},\ \bibinfo {pages} {15--38} (\bibinfo {year}
	{2015})}\BibitemShut {NoStop}%
\bibitem [{\citenamefont {Altman}\ and\ \citenamefont
	{Vosk}(2015)}]{altman_universal_2015}%
\BibitemOpen
\bibfield  {author} {\bibinfo {author} {\bibfnamefont {Ehud}\ \bibnamefont
		{Altman}}\ and\ \bibinfo {author} {\bibfnamefont {Ronen}\ \bibnamefont
		{Vosk}},\ }\bibfield  {title} {\enquote {\bibinfo {title} {Universal dynamics
			and renormalization in many body localized systems},}\ }\href {\doibase
	10.1146/annurev-conmatphys-031214-014701} {\bibfield  {journal} {\bibinfo
		{journal} {Annu. Rev. Condens. Matter}\ }\textbf {\bibinfo {volume} {6}},\
	\bibinfo {pages} {383--409} (\bibinfo {year} {2015})}\BibitemShut {NoStop}%
\bibitem [{\citenamefont {Schreiber}\ \emph {et~al.}(2015)\citenamefont
	{Schreiber}, \citenamefont {Hodgman}, \citenamefont {Bordia}, \citenamefont
	{L\"uschen}, \citenamefont {Fischer}, \citenamefont {Vosk}, \citenamefont
	{Altman}, \citenamefont {Schneider},\ and\ \citenamefont
	{Bloch}}]{Schreiber15}%
\BibitemOpen
\bibfield  {author} {\bibinfo {author} {\bibfnamefont {Michael}\ \bibnamefont
		{Schreiber}}, \bibinfo {author} {\bibfnamefont {Sean~S.}\ \bibnamefont
		{Hodgman}}, \bibinfo {author} {\bibfnamefont {Pranjal}\ \bibnamefont
		{Bordia}}, \bibinfo {author} {\bibfnamefont {Henrik~P.}\ \bibnamefont
		{L\"uschen}}, \bibinfo {author} {\bibfnamefont {Mark~H.}\ \bibnamefont
		{Fischer}}, \bibinfo {author} {\bibfnamefont {Ronen}\ \bibnamefont {Vosk}},
	\bibinfo {author} {\bibfnamefont {Ehud}\ \bibnamefont {Altman}}, \bibinfo
	{author} {\bibfnamefont {Ulrich}\ \bibnamefont {Schneider}}, \ and\ \bibinfo
	{author} {\bibfnamefont {Immanuel}\ \bibnamefont {Bloch}},\ }\bibfield
{title} {\enquote {\bibinfo {title} {Observation of many-body localization of
			interacting fermions in a quasirandom optical lattice},}\ }\href {\doibase
	10.1126/science.aaa7432} {\bibfield  {journal} {\bibinfo  {journal}
		{Science}\ }\textbf {\bibinfo {volume} {349}},\ \bibinfo {pages} {842--845}
	(\bibinfo {year} {2015})}\BibitemShut {NoStop}%
\bibitem [{\citenamefont {Kondov}\ \emph {et~al.}(2015)\citenamefont {Kondov},
	\citenamefont {McGehee}, \citenamefont {Xu},\ and\ \citenamefont
	{DeMarco}}]{kondov_Disorder_2015}%
\BibitemOpen
\bibfield  {author} {\bibinfo {author} {\bibfnamefont {S.~S.}\ \bibnamefont
		{Kondov}}, \bibinfo {author} {\bibfnamefont {W.~R.}\ \bibnamefont {McGehee}},
	\bibinfo {author} {\bibfnamefont {W.}~\bibnamefont {Xu}}, \ and\ \bibinfo
	{author} {\bibfnamefont {B.}~\bibnamefont {DeMarco}},\ }\bibfield  {title}
{\enquote {\bibinfo {title} {Disorder-induced localization in a strongly
			correlated atomic hubbard gas},}\ }\href {\doibase
	10.1103/PhysRevLett.114.083002} {\bibfield  {journal} {\bibinfo  {journal}
		{Phys. Rev. Lett.}\ }\textbf {\bibinfo {volume} {114}},\ \bibinfo {pages}
	{083002} (\bibinfo {year} {2015})}\BibitemShut {NoStop}%
\bibitem [{\citenamefont {Smith}\ \emph {et~al.}(2016)\citenamefont {Smith},
	\citenamefont {Lee}, \citenamefont {Richerme}, \citenamefont {Neyenhuis},
	\citenamefont {Hess}, \citenamefont {Hauke}, \citenamefont {Heyl},
	\citenamefont {Huse},\ and\ \citenamefont {Monroe}}]{smith2015}%
\BibitemOpen
\bibfield  {author} {\bibinfo {author} {\bibfnamefont {J.}~\bibnamefont
		{Smith}}, \bibinfo {author} {\bibfnamefont {A.}~\bibnamefont {Lee}}, \bibinfo
	{author} {\bibfnamefont {P.}~\bibnamefont {Richerme}}, \bibinfo {author}
	{\bibfnamefont {B.}~\bibnamefont {Neyenhuis}}, \bibinfo {author}
	{\bibfnamefont {P.~W.}\ \bibnamefont {Hess}}, \bibinfo {author}
	{\bibfnamefont {P.}~\bibnamefont {Hauke}}, \bibinfo {author} {\bibfnamefont
		{M.}~\bibnamefont {Heyl}}, \bibinfo {author} {\bibfnamefont {D.~A.}\
		\bibnamefont {Huse}}, \ and\ \bibinfo {author} {\bibfnamefont
		{C.}~\bibnamefont {Monroe}},\ }\bibfield  {title} {\enquote {\bibinfo {title}
		{Many-body localization in a quantum simulator with programmable random
			disorder},}\ }\href {\doibase 10.1038/nphys3783} {\bibfield  {journal}
	{\bibinfo  {journal} {Nat. Phys.}\ }\textbf {\bibinfo {volume} {12}},\
	\bibinfo {pages} {907--911} (\bibinfo {year} {2016})}\BibitemShut {NoStop}%
\bibitem [{\citenamefont {Bordia}\ \emph {et~al.}(2016)\citenamefont {Bordia},
	\citenamefont {L\"uschen}, \citenamefont {Hodgman}, \citenamefont
	{Schreiber}, \citenamefont {Bloch},\ and\ \citenamefont
	{Schneider}}]{Bordia16}%
\BibitemOpen
\bibfield  {author} {\bibinfo {author} {\bibfnamefont {Pranjal}\ \bibnamefont
		{Bordia}}, \bibinfo {author} {\bibfnamefont {Henrik~P.}\ \bibnamefont
		{L\"uschen}}, \bibinfo {author} {\bibfnamefont {Sean~S.}\ \bibnamefont
		{Hodgman}}, \bibinfo {author} {\bibfnamefont {Michael}\ \bibnamefont
		{Schreiber}}, \bibinfo {author} {\bibfnamefont {Immanuel}\ \bibnamefont
		{Bloch}}, \ and\ \bibinfo {author} {\bibfnamefont {Ulrich}\ \bibnamefont
		{Schneider}},\ }\bibfield  {title} {\enquote {\bibinfo {title} {Coupling
			identical one-dimensional many-body localized systems},}\ }\href {\doibase
	10.1103/PhysRevLett.116.140401} {\bibfield  {journal} {\bibinfo  {journal}
		{Phys. Rev. Lett.}\ }\textbf {\bibinfo {volume} {116}},\ \bibinfo {pages}
	{140401} (\bibinfo {year} {2016})}\BibitemShut {NoStop}%
\bibitem [{\citenamefont {Choi}\ \emph {et~al.}(2016)\citenamefont {Choi},
	\citenamefont {Hild}, \citenamefont {Zeiher}, \citenamefont {Schau{\ss}},
	\citenamefont {Rubio-Abadal}, \citenamefont {Yefsah}, \citenamefont
	{Khemani}, \citenamefont {Huse}, \citenamefont {Bloch},\ and\ \citenamefont
	{Gross}}]{choi_exploring_2016}%
\BibitemOpen
\bibfield  {author} {\bibinfo {author} {\bibfnamefont {Jae-yoon}\
		\bibnamefont {Choi}}, \bibinfo {author} {\bibfnamefont {Sebastian}\
		\bibnamefont {Hild}}, \bibinfo {author} {\bibfnamefont {Johannes}\
		\bibnamefont {Zeiher}}, \bibinfo {author} {\bibfnamefont {Peter}\
		\bibnamefont {Schau{\ss}}}, \bibinfo {author} {\bibfnamefont {Antonio}\
		\bibnamefont {Rubio-Abadal}}, \bibinfo {author} {\bibfnamefont {Tarik}\
		\bibnamefont {Yefsah}}, \bibinfo {author} {\bibfnamefont {Vedika}\
		\bibnamefont {Khemani}}, \bibinfo {author} {\bibfnamefont {David~A.}\
		\bibnamefont {Huse}}, \bibinfo {author} {\bibfnamefont {Immanuel}\
		\bibnamefont {Bloch}}, \ and\ \bibinfo {author} {\bibfnamefont {Christian}\
		\bibnamefont {Gross}},\ }\bibfield  {title} {\enquote {\bibinfo {title}
		{Exploring the many-body localization transition in two dimensions},}\ }\href
{\doibase 10.1126/science.aaf8834} {\bibfield  {journal} {\bibinfo  {journal}
		{Science}\ }\textbf {\bibinfo {volume} {352}},\ \bibinfo {pages} {1547--1552}
	(\bibinfo {year} {2016})}\BibitemShut {NoStop}%
\bibitem [{\citenamefont {Bordia}\ \emph {et~al.}()\citenamefont {Bordia},
	\citenamefont {L{\"u}schen}, \citenamefont {Schneider}, \citenamefont
	{Knap},\ and\ \citenamefont {Bloch}}]{bordia_periodically_2016}%
\BibitemOpen
\bibfield  {author} {\bibinfo {author} {\bibfnamefont {Pranjal}\ \bibnamefont
		{Bordia}}, \bibinfo {author} {\bibfnamefont {Henrik}\ \bibnamefont
		{L{\"u}schen}}, \bibinfo {author} {\bibfnamefont {Ulrich}\ \bibnamefont
		{Schneider}}, \bibinfo {author} {\bibfnamefont {Michael}\ \bibnamefont
		{Knap}}, \ and\ \bibinfo {author} {\bibfnamefont {Immanuel}\ \bibnamefont
		{Bloch}},\ }\bibfield  {title} {\enquote {\bibinfo {title} {Periodically
			driving a many-body localized quantum system},}\ }\href {\doibase
	10.1038/nphys4020} {\bibfield  {journal} {\bibinfo  {journal} {Nat. Phys.}\
	}\textbf {\bibinfo {volume} {13}},\ \bibinfo {pages} {460--464} (\bibinfo {year} {2017})}\BibitemShut
{NoStop}%
\bibitem [{\citenamefont {Basko}\ \emph {et~al.}(2007)\citenamefont {Basko},
	\citenamefont {Aleiner},\ and\ \citenamefont {Altshuler}}]{basko_expt}%
\BibitemOpen
\bibfield  {author} {\bibinfo {author} {\bibfnamefont {D.~M.}\ \bibnamefont
		{Basko}}, \bibinfo {author} {\bibfnamefont {I.~L.}\ \bibnamefont {Aleiner}},
	\ and\ \bibinfo {author} {\bibfnamefont {B.~L.}\ \bibnamefont {Altshuler}},\
}\bibfield  {title} {\enquote {\bibinfo {title} {Possible experimental
		manifestations of the many-body localization},}\ }\href {\doibase
10.1103/PhysRevB.76.052203} {\bibfield  {journal} {\bibinfo  {journal} {Phys.
		Rev. B}\ }\textbf {\bibinfo {volume} {76}},\ \bibinfo {pages} {052203}
(\bibinfo {year} {2007})}\BibitemShut {NoStop}%
\bibitem [{\citenamefont {Nandkishore}\ \emph {et~al.}(2014)\citenamefont
	{Nandkishore}, \citenamefont {Gopalakrishnan},\ and\ \citenamefont
	{Huse}}]{Nandkishore14}%
\BibitemOpen
\bibfield  {author} {\bibinfo {author} {\bibfnamefont {Rahul}\ \bibnamefont
		{Nandkishore}}, \bibinfo {author} {\bibfnamefont {Sarang}\ \bibnamefont
		{Gopalakrishnan}}, \ and\ \bibinfo {author} {\bibfnamefont {David~A.}\
		\bibnamefont {Huse}},\ }\bibfield  {title} {\enquote {\bibinfo {title}
		{Spectral features of a many-body-localized system weakly coupled to a
			bath},}\ }\href {\doibase 10.1103/PhysRevB.90.064203} {\bibfield  {journal}
	{\bibinfo  {journal} {Phys. Rev. B}\ }\textbf {\bibinfo {volume} {90}},\
	\bibinfo {pages} {064203} (\bibinfo {year} {2014})}\BibitemShut {NoStop}%
\bibitem [{\citenamefont {Gopalakrishnan}\ and\ \citenamefont
	{Nandkishore}(2014)}]{gopalakrishnan2014mean}%
\BibitemOpen
\bibfield  {author} {\bibinfo {author} {\bibfnamefont {Sarang}\ \bibnamefont
		{Gopalakrishnan}}\ and\ \bibinfo {author} {\bibfnamefont {Rahul}\
		\bibnamefont {Nandkishore}},\ }\bibfield  {title} {\enquote {\bibinfo {title}
		{Mean-field theory of nearly many-body localized metals},}\ }\href {\doibase
	10.1103/PhysRevB.90.224203} {\bibfield  {journal} {\bibinfo  {journal} {Phys.
			Rev. B}\ }\textbf {\bibinfo {volume} {90}},\ \bibinfo {pages} {224203}
	(\bibinfo {year} {2014})}\BibitemShut {NoStop}%
\bibitem [{\citenamefont {Johri}\ \emph {et~al.}(2015)\citenamefont {Johri},
	\citenamefont {Nandkishore},\ and\ \citenamefont {Bhatt}}]{johri2015many}%
\BibitemOpen
\bibfield  {author} {\bibinfo {author} {\bibfnamefont {Sonika}\ \bibnamefont
		{Johri}}, \bibinfo {author} {\bibfnamefont {Rahul}\ \bibnamefont
		{Nandkishore}}, \ and\ \bibinfo {author} {\bibfnamefont {RN}~\bibnamefont
		{Bhatt}},\ }\bibfield  {title} {\enquote {\bibinfo {title} {Many-body
			localization in imperfectly isolated quantum systems},}\ }\href {\doibase
	10.1103/PhysRevLett.114.117401} {\bibfield  {journal} {\bibinfo  {journal}
		{Phys. Rev. Lett.}\ }\textbf {\bibinfo {volume} {114}},\ \bibinfo {pages}
	{117401} (\bibinfo {year} {2015})}\BibitemShut {NoStop}%
\bibitem [{\citenamefont {Huse}\ \emph {et~al.}(2015)\citenamefont {Huse},
	\citenamefont {Nandkishore}, \citenamefont {Pietracaprina}, \citenamefont
	{Ros},\ and\ \citenamefont {Scardicchio}}]{huse2015localized}%
\BibitemOpen
\bibfield  {author} {\bibinfo {author} {\bibfnamefont {David~A}\ \bibnamefont
		{Huse}}, \bibinfo {author} {\bibfnamefont {Rahul}\ \bibnamefont
		{Nandkishore}}, \bibinfo {author} {\bibfnamefont {Francesca}\ \bibnamefont
		{Pietracaprina}}, \bibinfo {author} {\bibfnamefont {Valentina}\ \bibnamefont
		{Ros}}, \ and\ \bibinfo {author} {\bibfnamefont {Antonello}\ \bibnamefont
		{Scardicchio}},\ }\bibfield  {title} {\enquote {\bibinfo {title} {Localized
			systems coupled to small baths: From anderson to zeno},}\ }\href {\doibase
	10.1103/PhysRevB.92.014203} {\bibfield  {journal} {\bibinfo  {journal} {Phys.
			Rev. B}\ }\textbf {\bibinfo {volume} {92}},\ \bibinfo {pages} {014203}
	(\bibinfo {year} {2015})}\BibitemShut {NoStop}%
\bibitem [{\citenamefont {Hyatt}\ \emph {et~al.}(2017)\citenamefont {Hyatt},
	\citenamefont {Garrison}, \citenamefont {Potter},\ and\ \citenamefont
	{Bauer}}]{hyatt_many-body_2017}%
\BibitemOpen
\bibfield  {author} {\bibinfo {author} {\bibfnamefont {Katharine}\
		\bibnamefont {Hyatt}}, \bibinfo {author} {\bibfnamefont {James~R.}\
		\bibnamefont {Garrison}}, \bibinfo {author} {\bibfnamefont {Andrew~C.}\
		\bibnamefont {Potter}}, \ and\ \bibinfo {author} {\bibfnamefont {Bela}\
		\bibnamefont {Bauer}},\ }\bibfield  {title} {\enquote {\bibinfo {title}
		{Many-body localization in the presence of a small bath},}\ }\href {\doibase
	10.1103/PhysRevB.95.035132} {\bibfield  {journal} {\bibinfo  {journal} {Phys.
			Rev. B}\ }\textbf {\bibinfo {volume} {95}},\ \bibinfo {pages} {035132}
	(\bibinfo {year} {2017})}\BibitemShut {NoStop}%
\bibitem [{\citenamefont {Parameswaran}\ and\ \citenamefont
	{Gopalakrishnan}(2017)}]{parameswaran_spin-catalyzed_2017}%
\BibitemOpen
\bibfield  {author} {\bibinfo {author} {\bibfnamefont {S.~A.}\ \bibnamefont
		{Parameswaran}}\ and\ \bibinfo {author} {\bibfnamefont {S.}~\bibnamefont
		{Gopalakrishnan}},\ }\bibfield  {title} {\enquote {\bibinfo {title}
		{Spin-catalyzed hopping conductivity in disordered strongly interacting
			quantum wires},}\ }\href {\doibase 10.1103/PhysRevB.95.024201} {\bibfield
	{journal} {\bibinfo  {journal} {Phys. Rev. B}\ }\textbf {\bibinfo {volume}
		{95}},\ \bibinfo {pages} {024201} (\bibinfo {year} {2017})}\BibitemShut
{NoStop}%
\bibitem [{\citenamefont {Fischer}\ \emph {et~al.}(2016)\citenamefont
	{Fischer}, \citenamefont {Maksymenko},\ and\ \citenamefont
	{Altman}}]{fischer2016dynamics}%
\BibitemOpen
\bibfield  {author} {\bibinfo {author} {\bibfnamefont {Mark~H}\ \bibnamefont
		{Fischer}}, \bibinfo {author} {\bibfnamefont {Mykola}\ \bibnamefont
		{Maksymenko}}, \ and\ \bibinfo {author} {\bibfnamefont {Ehud}\ \bibnamefont
		{Altman}},\ }\bibfield  {title} {\enquote {\bibinfo {title} {Dynamics of a
			many-body-localized system coupled to a bath},}\ }\href {\doibase
	10.1103/PhysRevLett.116.160401} {\bibfield  {journal} {\bibinfo  {journal}
		{Phys. Rev. Lett.}\ }\textbf {\bibinfo {volume} {116}},\ \bibinfo {pages}
	{160401} (\bibinfo {year} {2016})}\BibitemShut {NoStop}%
\bibitem [{\citenamefont {Levi}\ \emph {et~al.}(2016)\citenamefont {Levi},
	\citenamefont {Heyl}, \citenamefont {Lesanovsky},\ and\ \citenamefont
	{Garrahan}}]{levi2016robustness}%
\BibitemOpen
\bibfield  {author} {\bibinfo {author} {\bibfnamefont {Emanuele}\
		\bibnamefont {Levi}}, \bibinfo {author} {\bibfnamefont {Markus}\ \bibnamefont
		{Heyl}}, \bibinfo {author} {\bibfnamefont {Igor}\ \bibnamefont {Lesanovsky}},
	\ and\ \bibinfo {author} {\bibfnamefont {Juan~P}\ \bibnamefont {Garrahan}},\
}\bibfield  {title} {\enquote {\bibinfo {title} {Robustness of many-body
		localization in the presence of dissipation},}\ }\href {\doibase
10.1103/PhysRevLett.116.237203} {\bibfield  {journal} {\bibinfo  {journal}
	{Phys. Rev. Lett.}\ }\textbf {\bibinfo {volume} {116}},\ \bibinfo {pages}
{237203} (\bibinfo {year} {2016})}\BibitemShut {NoStop}%
\bibitem [{\citenamefont {Medvedyeva}\ \emph {et~al.}(2016)\citenamefont
	{Medvedyeva}, \citenamefont {Prosen},\ and\ \citenamefont
	{{\v{Z}}nidari{\v{c}}}}]{medvedyeva2016influence}%
\BibitemOpen
\bibfield  {author} {\bibinfo {author} {\bibfnamefont {Mariya~V}\
		\bibnamefont {Medvedyeva}}, \bibinfo {author} {\bibfnamefont {Toma{\v{z}}}\
		\bibnamefont {Prosen}}, \ and\ \bibinfo {author} {\bibfnamefont {Marko}\
		\bibnamefont {{\v{Z}}nidari{\v{c}}}},\ }\bibfield  {title} {\enquote
	{\bibinfo {title} {Influence of dephasing on many-body localization},}\
}\href {\doibase 10.1103/PhysRevB.93.094205} {\bibfield  {journal} {\bibinfo
	{journal} {Phys. Rev. B}\ }\textbf {\bibinfo {volume} {93}},\ \bibinfo
{pages} {094205} (\bibinfo {year} {2016})}\BibitemShut {NoStop}%
\bibitem [{\citenamefont {Everest}\ \emph {et~al.}(2017)\citenamefont
	{Everest}, \citenamefont {Lesanovsky}, \citenamefont {Garrahan},\ and\
	\citenamefont {Levi}}]{everest_role_2017}%
\BibitemOpen
\bibfield  {author} {\bibinfo {author} {\bibfnamefont {Benjamin}\
		\bibnamefont {Everest}}, \bibinfo {author} {\bibfnamefont {Igor}\
		\bibnamefont {Lesanovsky}}, \bibinfo {author} {\bibfnamefont {Juan~P.}\
		\bibnamefont {Garrahan}}, \ and\ \bibinfo {author} {\bibfnamefont {Emanuele}\
		\bibnamefont {Levi}},\ }\bibfield  {title} {\enquote {\bibinfo {title} {Role
			of interactions in a dissipative many-body localized system},}\ }\href
{\doibase 10.1103/PhysRevB.95.024310} {\bibfield  {journal} {\bibinfo
		{journal} {Phys. Rev. B}\ }\textbf {\bibinfo {volume} {95}},\ \bibinfo
	{pages} {024310} (\bibinfo {year} {2017})}\BibitemShut {NoStop}%
\bibitem [{\citenamefont {Nandkishore}\ and\ \citenamefont
	{Gopalakrishnan}(2016)}]{nandkishore_general}%
\BibitemOpen
\bibfield  {author} {\bibinfo {author} {\bibfnamefont {Rahul}\ \bibnamefont
		{Nandkishore}}\ and\ \bibinfo {author} {\bibfnamefont {Sarang}\ \bibnamefont
		{Gopalakrishnan}},\ }\bibfield  {title} {\enquote {\bibinfo {title} {General
			theory of many body localized systems coupled to baths},}\ }\href {\doibase
	10.1002/andp.201600181} {\bibfield  {journal} {\bibinfo  {journal} {Ann.
			Phys.}\ } (\bibinfo {year} {2016}),\ 10.1002/andp.201600181}\BibitemShut
{NoStop}%
\bibitem [{\citenamefont {Bar~Lev}\ \emph {et~al.}(2015)\citenamefont
	{Bar~Lev}, \citenamefont {Cohen},\ and\ \citenamefont
	{Reichman}}]{BarLev_Absence_2015}%
\BibitemOpen
\bibfield  {author} {\bibinfo {author} {\bibfnamefont {Yevgeny}\ \bibnamefont
		{Bar~Lev}}, \bibinfo {author} {\bibfnamefont {Guy}\ \bibnamefont {Cohen}}, \
	and\ \bibinfo {author} {\bibfnamefont {David~R.}\ \bibnamefont {Reichman}},\
}\bibfield  {title} {\enquote {\bibinfo {title} {Absence of diffusion in an
		interacting system of spinless fermions on a one-dimensional disordered
		lattice},}\ }\href {\doibase 10.1103/PhysRevLett.114.100601} {\bibfield
{journal} {\bibinfo  {journal} {Phys. Rev. Lett.}\ }\textbf {\bibinfo
	{volume} {114}},\ \bibinfo {pages} {100601} (\bibinfo {year}
{2015})}\BibitemShut {NoStop}%
\bibitem [{\citenamefont {Agarwal}\ \emph {et~al.}(2015)\citenamefont
	{Agarwal}, \citenamefont {Gopalakrishnan}, \citenamefont {Knap},
	\citenamefont {M{\"u}ller},\ and\ \citenamefont
	{Demler}}]{agarwal_anomalous_2015}%
\BibitemOpen
\bibfield  {author} {\bibinfo {author} {\bibfnamefont {Kartiek}\ \bibnamefont
		{Agarwal}}, \bibinfo {author} {\bibfnamefont {Sarang}\ \bibnamefont
		{Gopalakrishnan}}, \bibinfo {author} {\bibfnamefont {Michael}\ \bibnamefont
		{Knap}}, \bibinfo {author} {\bibfnamefont {Markus}\ \bibnamefont
		{M{\"u}ller}}, \ and\ \bibinfo {author} {\bibfnamefont {Eugene}\ \bibnamefont
		{Demler}},\ }\bibfield  {title} {\enquote {\bibinfo {title} {Anomalous
			diffusion and griffiths effects near the many-body localization
			transition},}\ }\href {\doibase 10.1103/PhysRevLett.114.160401} {\bibfield
	{journal} {\bibinfo  {journal} {Phys. Rev. Lett.}\ }\textbf {\bibinfo
		{volume} {114}},\ \bibinfo {pages} {160401} (\bibinfo {year}
	{2015})}\BibitemShut {NoStop}%
\bibitem [{\citenamefont {Vosk}\ \emph {et~al.}(2015)\citenamefont {Vosk},
	\citenamefont {Huse},\ and\ \citenamefont {Altman}}]{Vosk_Theory_2015}%
\BibitemOpen
\bibfield  {author} {\bibinfo {author} {\bibfnamefont {Ronen}\ \bibnamefont
		{Vosk}}, \bibinfo {author} {\bibfnamefont {David~A.}\ \bibnamefont {Huse}}, \
	and\ \bibinfo {author} {\bibfnamefont {Ehud}\ \bibnamefont {Altman}},\
}\bibfield  {title} {\enquote {\bibinfo {title} {Theory of the many-body
		localization transition in one-dimensional systems},}\ }\href {\doibase
10.1103/PhysRevX.5.031032} {\bibfield  {journal} {\bibinfo  {journal} {Phys.
		Rev. X}\ }\textbf {\bibinfo {volume} {5}},\ \bibinfo {pages} {031032}
(\bibinfo {year} {2015})}\BibitemShut {NoStop}%
\bibitem [{\citenamefont {Potter}\ \emph {et~al.}(2015)\citenamefont {Potter},
	\citenamefont {Vasseur},\ and\ \citenamefont
	{Parameswaran}}]{Potter_Universal_2015}%
\BibitemOpen
\bibfield  {author} {\bibinfo {author} {\bibfnamefont {Andrew~C.}\
		\bibnamefont {Potter}}, \bibinfo {author} {\bibfnamefont {Romain}\
		\bibnamefont {Vasseur}}, \ and\ \bibinfo {author} {\bibfnamefont {S.~A.}\
		\bibnamefont {Parameswaran}},\ }\bibfield  {title} {\enquote {\bibinfo
		{title} {Universal properties of many-body delocalization transitions},}\
}\href {\doibase 10.1103/PhysRevX.5.031033} {\bibfield  {journal} {\bibinfo
	{journal} {Phys. Rev. X}\ }\textbf {\bibinfo {volume} {5}},\ \bibinfo {pages}
{031033} (\bibinfo {year} {2015})}\BibitemShut {NoStop}%
\bibitem [{\citenamefont {Luitz}\ \emph {et~al.}(2016)\citenamefont {Luitz},
	\citenamefont {Laflorencie},\ and\ \citenamefont {Alet}}]{luitz2016}%
\BibitemOpen
\bibfield  {author} {\bibinfo {author} {\bibfnamefont {David~J.}\
		\bibnamefont {Luitz}}, \bibinfo {author} {\bibfnamefont {Nicolas}\
		\bibnamefont {Laflorencie}}, \ and\ \bibinfo {author} {\bibfnamefont
		{Fabien}\ \bibnamefont {Alet}},\ }\bibfield  {title} {\enquote {\bibinfo
		{title} {Extended slow dynamical regime close to the many-body localization
			transition},}\ }\href {\doibase 10.1103/PhysRevB.93.060201} {\bibfield
	{journal} {\bibinfo  {journal} {Phys. Rev. B}\ }\textbf {\bibinfo {volume}
		{93}},\ \bibinfo {pages} {060201} (\bibinfo {year} {2016})}\BibitemShut
{NoStop}%
\bibitem [{\citenamefont {Gopalakrishnan}\ \emph {et~al.}(2016)\citenamefont
	{Gopalakrishnan}, \citenamefont {Agarwal}, \citenamefont {Demler},
	\citenamefont {Huse},\ and\ \citenamefont
	{Knap}}]{gopalakrishnan_griffiths_2016}%
\BibitemOpen
\bibfield  {author} {\bibinfo {author} {\bibfnamefont {Sarang}\ \bibnamefont
		{Gopalakrishnan}}, \bibinfo {author} {\bibfnamefont {Kartiek}\ \bibnamefont
		{Agarwal}}, \bibinfo {author} {\bibfnamefont {Eugene~A.}\ \bibnamefont
		{Demler}}, \bibinfo {author} {\bibfnamefont {David~A.}\ \bibnamefont {Huse}},
	\ and\ \bibinfo {author} {\bibfnamefont {Michael}\ \bibnamefont {Knap}},\
}\bibfield  {title} {\enquote {\bibinfo {title} {Griffiths effects and slow
		dynamics in nearly many-body localized systems},}\ }\href {\doibase
10.1103/PhysRevB.93.134206} {\bibfield  {journal} {\bibinfo  {journal} {Phys.
		Rev. B}\ }\textbf {\bibinfo {volume} {93}},\ \bibinfo {pages} {134206}
(\bibinfo {year} {2016})}\BibitemShut {NoStop}%
\bibitem [{\citenamefont {Agarwal}\ \emph {et~al.}(2017)\citenamefont
	{Agarwal}, \citenamefont {Altman}, \citenamefont {Demler}, \citenamefont
	{Gopalakrishnan}, \citenamefont {Huse},\ and\ \citenamefont
	{Knap}}]{agarwal_rr_16}%
\BibitemOpen
\bibfield  {author} {\bibinfo {author} {\bibfnamefont {Kartiek}\ \bibnamefont
		{Agarwal}}, \bibinfo {author} {\bibfnamefont {Ehud}\ \bibnamefont {Altman}},
	\bibinfo {author} {\bibfnamefont {Eugene}\ \bibnamefont {Demler}}, \bibinfo
	{author} {\bibfnamefont {Sarang}\ \bibnamefont {Gopalakrishnan}}, \bibinfo
	{author} {\bibfnamefont {David~A.}\ \bibnamefont {Huse}}, \ and\ \bibinfo
	{author} {\bibfnamefont {Michael}\ \bibnamefont {Knap}},\ }\bibfield  {title}
{\enquote {\bibinfo {title} {Rare region effects and dynamics near the
			many-body localization transition},}\ }\href {\doibase
	10.1002/andp.201600326} {\bibfield  {journal} {\bibinfo  {journal} {Ann.
			Phys.}\ }\textbf {\bibinfo {volume} {529}},\ \bibinfo {pages} {1600326}
	(\bibinfo {year} {2017})}\BibitemShut {NoStop}%
\bibitem [{\citenamefont {L\"uschen}\ \emph {et~al.}(2016)\citenamefont
	{L\"uschen}, \citenamefont {Bordia}, \citenamefont {Scherg}, \citenamefont
	{Alet}, \citenamefont {Altman}, \citenamefont {Schneider},\ and\
	\citenamefont {Bloch}}]{lueschencrit16}%
\BibitemOpen
\bibfield  {author} {\bibinfo {author} {\bibfnamefont {Henrik~P.}\
		\bibnamefont {L\"uschen}}, \bibinfo {author} {\bibfnamefont {Pranjal}\
		\bibnamefont {Bordia}}, \bibinfo {author} {\bibfnamefont {Sebastian}\
		\bibnamefont {Scherg}}, \bibinfo {author} {\bibfnamefont {Fabien}\
		\bibnamefont {Alet}}, \bibinfo {author} {\bibfnamefont {Ehud}\ \bibnamefont
		{Altman}}, \bibinfo {author} {\bibfnamefont {Ulrich}\ \bibnamefont
		{Schneider}}, \ and\ \bibinfo {author} {\bibfnamefont {Immanuel}\
		\bibnamefont {Bloch}},\ }\bibfield  {title} {\enquote {\bibinfo {title}
		{Evidence for {G}riffiths-type dynamics near the many-body localization
			transition in quasi-periodic systems},}\ }\href@noop {} {\  (\bibinfo {year}
	{2016})},\ \Eprint {http://arxiv.org/abs/arXiv:1612.07173} {arXiv:1612.07173}
\BibitemShut {NoStop}%
\bibitem [{\citenamefont {Bordia}\ \emph {et~al.}(2017)\citenamefont {Bordia},
	\citenamefont {L\"uschen}, \citenamefont {Scherg}, \citenamefont
	{Gopalakrishnan}, \citenamefont {Knap}, \citenamefont {Schneider},\ and\
	\citenamefont {Bloch}}]{bordia2d17}%
\BibitemOpen
\bibfield  {author} {\bibinfo {author} {\bibfnamefont {Pranjal}\ \bibnamefont
		{Bordia}}, \bibinfo {author} {\bibfnamefont {Henrik~P.}\ \bibnamefont
		{L\"uschen}}, \bibinfo {author} {\bibfnamefont {Sebastian}\ \bibnamefont
		{Scherg}}, \bibinfo {author} {\bibfnamefont {Sarang}\ \bibnamefont
		{Gopalakrishnan}}, \bibinfo {author} {\bibfnamefont {Michael}\ \bibnamefont
		{Knap}}, \bibinfo {author} {\bibfnamefont {Ulrich}\ \bibnamefont
		{Schneider}}, \ and\ \bibinfo {author} {\bibfnamefont {Immanuel}\
		\bibnamefont {Bloch}},\ }\bibfield  {title} {\enquote {\bibinfo {title}
		{Probing slow relaxation and many-body localization in two-dimensional
			quasi-periodic systems},}\ }\href@noop {} {\  (\bibinfo {year} {2017})},\
\Eprint {http://arxiv.org/abs/arXiv:1704.03063} {arXiv:1704.03063}
\BibitemShut {NoStop}%
\bibitem [{\citenamefont {Gopalakrishnan}\ \emph {et~al.}(2015)\citenamefont
	{Gopalakrishnan}, \citenamefont {M\"uller}, \citenamefont {Khemani},
	\citenamefont {Knap}, \citenamefont {Demler},\ and\ \citenamefont
	{Huse}}]{Gopalakrishnan15}%
\BibitemOpen
\bibfield  {author} {\bibinfo {author} {\bibfnamefont {Sarang}\ \bibnamefont
		{Gopalakrishnan}}, \bibinfo {author} {\bibfnamefont {Markus}\ \bibnamefont
		{M\"uller}}, \bibinfo {author} {\bibfnamefont {Vedika}\ \bibnamefont
		{Khemani}}, \bibinfo {author} {\bibfnamefont {Michael}\ \bibnamefont {Knap}},
	\bibinfo {author} {\bibfnamefont {Eugene}\ \bibnamefont {Demler}}, \ and\
	\bibinfo {author} {\bibfnamefont {David~A.}\ \bibnamefont {Huse}},\
}\bibfield  {title} {\enquote {\bibinfo {title} {Low-frequency conductivity
		in many-body localized systems},}\ }\href {\doibase
10.1103/PhysRevB.92.104202} {\bibfield  {journal} {\bibinfo  {journal} {Phys.
		Rev. B}\ }\textbf {\bibinfo {volume} {92}},\ \bibinfo {pages} {104202}
(\bibinfo {year} {2015})}\BibitemShut {NoStop}%
\bibitem [{\citenamefont {Pekker}\ \emph {et~al.}(2016)\citenamefont {Pekker},
	\citenamefont {Clark}, \citenamefont {Oganesyan},\ and\ \citenamefont
	{Refael}}]{pekker_2016}%
\BibitemOpen
\bibfield  {author} {\bibinfo {author} {\bibfnamefont {David}\ \bibnamefont
		{Pekker}}, \bibinfo {author} {\bibfnamefont {Bryan~K}\ \bibnamefont {Clark}},
	\bibinfo {author} {\bibfnamefont {Vadim}\ \bibnamefont {Oganesyan}}, \ and\
	\bibinfo {author} {\bibfnamefont {Gil}\ \bibnamefont {Refael}},\ }\bibfield
{title} {\enquote {\bibinfo {title} {Fixed points of wegner-wilson flows and
			many-body localization},}\ }\href {http://arxiv.org/abs/1607.07884}
{\bibfield  {journal} {\bibinfo  {journal} {arXiv:1607.07884}\ } (\bibinfo
	{year} {2016})}\BibitemShut {NoStop}%
\bibitem [{\citenamefont {Serbyn}\ \emph {et~al.}(2014)\citenamefont {Serbyn},
	\citenamefont {Knap}, \citenamefont {Gopalakrishnan}, \citenamefont
	{Papi{\'c}}, \citenamefont {Yao}, \citenamefont {Laumann}, \citenamefont
	{Abanin}, \citenamefont {Lukin},\ and\ \citenamefont
	{Demler}}]{serbyn_interferometric_2014}%
\BibitemOpen
\bibfield  {author} {\bibinfo {author} {\bibfnamefont {M.}~\bibnamefont
		{Serbyn}}, \bibinfo {author} {\bibfnamefont {M.}~\bibnamefont {Knap}},
	\bibinfo {author} {\bibfnamefont {S.}~\bibnamefont {Gopalakrishnan}},
	\bibinfo {author} {\bibfnamefont {Z.}~\bibnamefont {Papi{\'c}}}, \bibinfo
	{author} {\bibfnamefont {N. Y.}\ \bibnamefont {Yao}}, \bibinfo {author}
	{\bibfnamefont {C. R.}\ \bibnamefont {Laumann}}, \bibinfo {author}
	{\bibfnamefont {D. A.}\ \bibnamefont {Abanin}}, \bibinfo {author}
	{\bibfnamefont {M. D.}\ \bibnamefont {Lukin}}, \ and\ \bibinfo {author}
	{\bibfnamefont {E. A.}\ \bibnamefont {Demler}},\ }\bibfield  {title}
{\enquote {\bibinfo {title} {Interferometric probes of many-body
			localization},}\ }\href {\doibase 10.1103/PhysRevLett.113.147204} {\bibfield
	{journal} {\bibinfo  {journal} {Phys. Rev. Lett.}\ }\textbf {\bibinfo
		{volume} {113}},\ \bibinfo {pages} {147204} (\bibinfo {year}
	{2014})}\BibitemShut {NoStop}%
\bibitem [{\citenamefont {Bouchaud}\ and\ \citenamefont
	{Georges}(1990)}]{bouchaud_anomalous_1990}%
\BibitemOpen
\bibfield  {author} {\bibinfo {author} {\bibfnamefont {Jean-Philippe}\
		\bibnamefont {Bouchaud}}\ and\ \bibinfo {author} {\bibfnamefont {Antoine}\
		\bibnamefont {Georges}},\ }\bibfield  {title} {\enquote {\bibinfo {title}
		{Anomalous diffusion in disordered media: Statistical mechanisms, models and
			physical applications},}\ }\href {\doibase 10.1016/0370-1573(90)90099-N}
{\bibfield  {journal} {\bibinfo  {journal} {Phys. Rep.}\ }\textbf {\bibinfo
		{volume} {195}},\ \bibinfo {pages} {127--293} (\bibinfo {year}
	{1990})}\BibitemShut {NoStop}%
\bibitem [{\citenamefont {Crow}\ and\ \citenamefont {Joynt}(2014)}]{crow_2014}%
\BibitemOpen
\bibfield  {author} {\bibinfo {author} {\bibfnamefont {Daniel}\ \bibnamefont
		{Crow}}\ and\ \bibinfo {author} {\bibfnamefont {Robert}\ \bibnamefont
		{Joynt}},\ }\bibfield  {title} {\enquote {\bibinfo {title} {Classical
			simulation of quantum dephasing and depolarizing noise},}\ }\href {\doibase
	10.1103/PhysRevA.89.042123} {\bibfield  {journal} {\bibinfo  {journal} {Phys.
			Rev. A}\ }\textbf {\bibinfo {volume} {89}},\ \bibinfo {pages} {042123}
	(\bibinfo {year} {2014})}\BibitemShut {NoStop}%
\bibitem [{\citenamefont {Gardiner}\ \emph {et~al.}(1985)\citenamefont
	{Gardiner} \emph {et~al.}}]{gardiner1985handbook}%
\BibitemOpen
\bibfield  {author} {\bibinfo {author} {\bibfnamefont {Crispin~W}\
		\bibnamefont {Gardiner}} \emph {et~al.},\ }\href@noop {} {\emph {\bibinfo
		{title} {Handbook of stochastic methods}}},\ Vol.~\bibinfo {volume} {3}\
(\bibinfo  {publisher} {Springer Berlin},\ \bibinfo {year}
{1985})\BibitemShut {NoStop}%
\bibitem [{\citenamefont {Amir}\ \emph {et~al.}(2009)\citenamefont {Amir},
	\citenamefont {Lahini},\ and\ \citenamefont {Perets}}]{amir_classical_2009}%
\BibitemOpen
\bibfield  {author} {\bibinfo {author} {\bibfnamefont {Ariel}\ \bibnamefont
		{Amir}}, \bibinfo {author} {\bibfnamefont {Yoav}\ \bibnamefont {Lahini}}, \
	and\ \bibinfo {author} {\bibfnamefont {Hagai~B.}\ \bibnamefont {Perets}},\
}\bibfield  {title} {\enquote {\bibinfo {title} {Classical diffusion of a
		quantum particle in a noisy environment},}\ }\href {\doibase
10.1103/PhysRevE.79.050105} {\bibfield  {journal} {\bibinfo  {journal} {Phys.
		Rev. E}\ }\textbf {\bibinfo {volume} {79}},\ \bibinfo {pages} {050105}
(\bibinfo {year} {2009})}\BibitemShut {NoStop}%
\bibitem [{sup()}]{supp}%
\BibitemOpen
\href@noop {} {}\bibinfo {note} {See supplementary online
	material}\BibitemShut {NoStop}%
\bibitem [{\citenamefont {{\v Z}nidari{\v c}}\ and\ \citenamefont
	{Horvat}(2013)}]{znidaric_transport_2013}%
\BibitemOpen
\bibfield  {author} {\bibinfo {author} {\bibfnamefont {Marko}\ \bibnamefont
		{{\v Z}nidari{\v c}}}\ and\ \bibinfo {author} {\bibfnamefont {Martin}\
		\bibnamefont {Horvat}},\ }\bibfield  {title} {\enquote {\bibinfo {title}
		{Transport in a disordered tight-binding chain with dephasing},}\ }\href
{\doibase 10.1140/epjb/e2012-30730-9} {\bibfield  {journal} {\bibinfo
		{journal} {Eur. Phys. J. B}\ }\textbf {\bibinfo {volume} {86}},\ \bibinfo
	{pages} {67} (\bibinfo {year} {2013})}\BibitemShut {NoStop}%
\bibitem [{\citenamefont {Minwalla}\ \emph {et~al.}(2000)\citenamefont
	{Minwalla}, \citenamefont {Van~Raamsdonk},\ and\ \citenamefont
	{Seiberg}}]{minwalla2000noncommutative}%
\BibitemOpen
\bibfield  {author} {\bibinfo {author} {\bibfnamefont {Shiraz}\ \bibnamefont
		{Minwalla}}, \bibinfo {author} {\bibfnamefont {Mark}\ \bibnamefont
		{Van~Raamsdonk}}, \ and\ \bibinfo {author} {\bibfnamefont {Nathan}\
		\bibnamefont {Seiberg}},\ }\bibfield  {title} {\enquote {\bibinfo {title}
		{Noncommutative perturbative dynamics},}\ }\href {\doibase
	10.1088/1126-6708/2000/02/020} {\bibfield  {journal} {\bibinfo  {journal} {J.
			High Energy Phys.}\ }\textbf {\bibinfo {volume} {2000}},\ \bibinfo {pages}
	{020} (\bibinfo {year} {2000})}\BibitemShut {NoStop}%
\bibitem [{\citenamefont {Lucioni}\ \emph {et~al.}(2011)\citenamefont
	{Lucioni}, \citenamefont {Deissler}, \citenamefont {Tanzi}, \citenamefont
	{Roati}, \citenamefont {Zaccanti}, \citenamefont {Modugno}, \citenamefont
	{Larcher}, \citenamefont {Dalfovo}, \citenamefont {Inguscio},\ and\
	\citenamefont {Modugno}}]{lucioni_observation_2011}%
\BibitemOpen
\bibfield  {author} {\bibinfo {author} {\bibfnamefont {E.}~\bibnamefont
		{Lucioni}}, \bibinfo {author} {\bibfnamefont {B.}~\bibnamefont {Deissler}},
	\bibinfo {author} {\bibfnamefont {L.}~\bibnamefont {Tanzi}}, \bibinfo
	{author} {\bibfnamefont {G.}~\bibnamefont {Roati}}, \bibinfo {author}
	{\bibfnamefont {M.}~\bibnamefont {Zaccanti}}, \bibinfo {author}
	{\bibfnamefont {M.}~\bibnamefont {Modugno}}, \bibinfo {author} {\bibfnamefont
		{M.}~\bibnamefont {Larcher}}, \bibinfo {author} {\bibfnamefont
		{F.}~\bibnamefont {Dalfovo}}, \bibinfo {author} {\bibfnamefont
		{M.}~\bibnamefont {Inguscio}}, \ and\ \bibinfo {author} {\bibfnamefont
		{G.}~\bibnamefont {Modugno}},\ }\bibfield  {title} {\enquote {\bibinfo
		{title} {Observation of {Subdiffusion} in a {Disordered} {Interacting}
			{System}},}\ }\href {\doibase 10.1103/PhysRevLett.106.230403} {\bibfield
	{journal} {\bibinfo  {journal} {Phys. Rev. Lett.}\ }\textbf {\bibinfo
		{volume} {106}},\ \bibinfo {pages} {230403} (\bibinfo {year}
	{2011})}\BibitemShut {NoStop}%
\bibitem [{\citenamefont
	{Shepelyansky}(1993)}]{shepelyansky_delocalization_1993}%
\BibitemOpen
\bibfield  {author} {\bibinfo {author} {\bibfnamefont {D.~L.}\ \bibnamefont
		{Shepelyansky}},\ }\bibfield  {title} {\enquote {\bibinfo {title}
		{Delocalization of quantum chaos by weak nonlinearity},}\ }\href {\doibase
	10.1103/PhysRevLett.70.1787} {\bibfield  {journal} {\bibinfo  {journal}
		{Phys. Rev. Lett.}\ }\textbf {\bibinfo {volume} {70}},\ \bibinfo {pages}
	{1787--1790} (\bibinfo {year} {1993})}\BibitemShut {NoStop}%
\bibitem [{\citenamefont {Flach}\ \emph {et~al.}(2009)\citenamefont {Flach},
	\citenamefont {Krimer},\ and\ \citenamefont {Skokos}}]{flach_universal_2009}%
\BibitemOpen
\bibfield  {author} {\bibinfo {author} {\bibfnamefont {S.}~\bibnamefont
		{Flach}}, \bibinfo {author} {\bibfnamefont {D.~O.}\ \bibnamefont {Krimer}}, \
	and\ \bibinfo {author} {\bibfnamefont {Ch.}\ \bibnamefont {Skokos}},\
}\bibfield  {title} {\enquote {\bibinfo {title} {Universal {Spreading} of
		{Wave} {Packets} in {Disordered} {Nonlinear} {Systems}},}\ }\href {\doibase
10.1103/PhysRevLett.102.024101} {\bibfield  {journal} {\bibinfo  {journal}
	{Phys. Rev. Lett.}\ }\textbf {\bibinfo {volume} {102}},\ \bibinfo {pages}
{024101} (\bibinfo {year} {2009})}\BibitemShut {NoStop}%
\end{thebibliography}
\end{document}